%% file: manuscript.tex
\documentclass[preprint2]{aastex631}
\usepackage{amsmath}
\usepackage{comment}
\usepackage{gensymb}

\newcommand{\exofasttwo}{{\tt EXOFASTv2}}

\begin{document}
\nolinenumbers

\title{JWST Thermal Emission of the Terrestrial Exoplanet GJ\,1132b}

\author[0000-0002-6215-5425]{Qiao Xue}
\affiliation{Department of Astronomy \& Astrophysics, University of Chicago, Chicago, IL, USA}

\author[0000-0003-4733-6532]{Jacob L.\ Bean}
\affiliation{Department of Astronomy \& Astrophysics, University of Chicago, Chicago, IL, USA}

\author[0000-0002-0659-1783]{Michael Zhang}
\affiliation{Department of Astronomy \& Astrophysics, University of Chicago, Chicago, IL, USA}

\author[0009-0000-8049-3797]{Alexandra Mahajan}
\affiliation{Center for Astrophysics \textbar \ Harvard \& Smithsonian, 60 Garden St, Cambridge, MA 02138, USA}

\author[0000-0003-2775-653X]{Jegug Ih}
\affiliation{Department of Astronomy, University of Maryland, College Park, MD, USA}

\author[0000-0003-3773-5142]{Jason D.\ Eastman}
\affiliation{Center for Astrophysics \textbar \ Harvard \& Smithsonian, 60 Garden St, Cambridge, MA 02138, USA}

\author[0000-0003-2279-4131]{Jonathan Lunine}
\affiliation{Department of Astronomy, Cornell University, Ithaca, NY, USA}

\author[0000-0003-4241-7413]{Megan Weiner Mansfield}
\affiliation{Steward Observatory, University of Arizona, Tucson, AZ, USA}
\affiliation{NHFP Sagan Fellow}

\author[0000-0002-0508-857X]{Brandon Park Coy}
\affiliation{Department of the Geophysical Sciences, University of Chicago, Chicago, IL, 60637, USA}

\author[0000-0002-1337-9051]{Eliza M.-R. Kempton}
\affiliation{Department of Astronomy, University of Maryland, College Park, MD, USA}

\author[0000-0002-9076-6901]{Daniel Koll}
\affiliation{Peking University, Beijing, People's Republic of China}

\author[0000-0002-1426-1186]{Edwin Kite}
\affiliation{Department of the Geophysical Sciences, University of Chicago, Chicago, IL, 60637, USA}



\begin{abstract}
We present thermal emission measurements of GJ\,1132b spanning 5 -- 12\,$\mu m$ obtained with the Mid-Infrared Instrument Low-Resolution Spectrometer (MIRI/LRS) on the James Webb Space Telescope (JWST). GJ\,1132b is an M-dwarf rocky planet with $T_{\mathrm{eq}}\,=\,584$\,K and an orbital period of 1.6 days. We measure a white-light secondary eclipse depth of $140 \pm 17$\,ppm, which corresponds to a dayside brightness temperature of $T_{p,\mathrm{dayside}} = 709\pm 31$\,K using improved star and planet parameters. This measured temperature is only 1$\sigma$ below the maximum possible dayside temperature of a bare rock (i.e., assuming a zero albedo planet with no heat redistribution, $T_{\mathrm{max}}$ = 746$^{+14}_{-11}$\,K). The emission spectrum is consistent with a featureless blackbody, which agrees with a wide range of possible surface compositions. By comparing forward models to the dayside emission spectrum, we rule out Earth-thickness (P 
$\sim$ 1\,bar) atmospheres with at least 1\% H$_2$O, atmospheres of any modeled thickness (10$^{-4}$ -- 10$^2$\,bar) that contain at least 1\% CO$_2$, and thick, Venus-like atmospheres ($P\,\gtrsim$\,100\,bar) with at least 1\,ppm CO$_2$ or H$_2$O. We therefore conclude that GJ\,1132b likely does not have a significant atmosphere. This finding supports the concept of a universal ``Cosmic Shoreline'' given the high level of bolometric and XUV irradiation received by the planet.


\end{abstract}


\section{Introduction} \label{sec:intro}
One of the most exciting questions in astronomy right now is whether small rocky planets orbiting M dwarfs can host atmospheres, which is directly related to their habitability. About 61\% of the stars within 10\,pc around us are M stars \citep{henry_solar_2018, reyle_10_2021}, and it has become clear from the Kepler mission \citep{mulders_stellar-mass-dependent_2015, dressing_occurrence_2015} and radial velocity surveys \citep{sabotta_carmenes_2021} that these stars preferentially host rocky planets. More importantly, their lower luminosity and smaller radius relative to Sun-like stars open a window for studying the atmospheres of transiting habitable M-dwarf planets using existing facilities. 

However, M dwarfs are also known to be very active \citep{west_constraining_2008}. UV emission during the pre-main sequence stage of M stars drives water loss and O$_2$ buildup \citep{luger_extreme_2015}. On the other side, powerful flares and strong UV radiation may destroy the atmosphere of the planets through photoevaporation or stellar wind erosion \citep{zendejas_atmospheric_2010, do_amaral_contribution_2022,affolter_planetary_2023}. 

No previous measurements have found strong evidence for atmospheres on rocky M-dwarf planets, and recent JWST results are highly constraining non-detections or at most inconclusive. Mid-Infrared Instrument (MIRI) emission observations of three M-dwarf rocky planets, TRAPPIST-1\,b, c (both 15\,$\mu$m filter photometry), and GJ\,367b (5 -- 12\,$\mu$m low resolution spectroscopy), have shown no evidence of substantial atmospheres \citep{greene_thermal_2023,zieba_no_2023,Ih_T1b_2023,lincowski2023,zhang_gj_2024}. On the other hand, with transmission spectroscopy, the results are unclear either due to possible stellar contamination (Gl\,486b: \citealt{moran_high_2023}; TRAPPIST-1\,b: \citealt{lim_atmospheric_2023}; GJ\,1132b: \citealt{may_double_2023}; LHS\,1140b: \citealt{cadieux24_lhs1140}) or degeneracy of different scenarios that could cause featureless spectra, e.g., a high mean molecular weight atmosphere, a high-altitude cloud atmosphere, no or very thin atmosphere (LHS\,475b, \citealt{lustig-yaeger_jwst_2023}; GJ\,341b, \citealt{kirk_jwstnircam_2024}; TOI-836b, \citealt{alderson_jwst_2024}).

GJ\,1132b, first detected by the MEarth-South telescope array \citep{berta-thompson_rocky_2015}, is a super-Earth with a relatively low equilibrium temperature $T_\mathrm{eq}\sim584$\,K  orbiting a nearby M4-type dwarf star. Previous efforts to detect an atmosphere on GJ\,1132b using transmission spectroscopy with HST/WFC3 \citep{swain_detection_2021,mugnai_ares_2021,libby-roberts_featureless_2022} and JWST NIRSpec G395H \citep{may_double_2023} have all been inconclusive about the presence of secondary atmospheres. 
GJ\,1132 is considered to be a weakly active M-star with slow rotation velocity $<$2\,km\,s$^{-1}$, very long rotation period $>100$\,days and age $>5$\,Gyr \citep{berta-thompson_rocky_2015, bonfils_radial_2018}. GJ\,1132b is considered to be one of the most suitable rocky planets for thermal emission measurements with JWST according to the emission spectroscopy metric (ESM $\sim$ 10) of \citet{kempton_framework_2018}\footnote{\url{https://tess.mit.edu/science/tess-acwg/}}. 

In this study, we use the technique of secondary eclipse thermal emission to probe the rocky planet GJ\,1132b for the presence of an atmosphere. This technique was originally developed by \citet{koll_identifying_2019} and \citet{mansfield_identifying_2019} building off the ideas of \citet{deming_discovery_2009} and \citet{selsis_thermal_2011}. In addition to the recent JWST results described above, this technique has previously been applied using Spitzer Space Telescope observations (LHS\,3844b: \citealt{kreidberg_absence_2019, Whittaker_lhs3844_2022}; GJ\,1252b: \citealt{crossfield_gj_2022}).  However, prior Spitzer observations of GJ\,1132b were not able to detect the secondary eclipse due to instrumental limits \citep{dittmann_search_2017}.

In this work, we present JWST/MIRI thermal emission measurements of GJ\,1132b. We describe the observation details in \S \ref{sec:observation}, and provide an overview of the data reduction in \S \ref{sec:data_reduction}. Refined stellar and planet parameter measurements are described in \ref{sec:stellar_parameters}. We present the results in \S \ref{sec:interpretation} and discussion in \S \ref{sec:discussion}.

\section{Observation} \label{sec:observation}
We observed one secondary eclipse of GJ\,1132b using JWST MIRI on July 1, 2023 (program GTO 1274, J.~Lunine PI), with an exposure length of 4.5 hours. A single eclipse was planned because \citet{koll_identifying_2019} showed that this would give sufficient signal-to-noise to be highly constraining for potential atmospheres. The observation was taken in the Low Resolution Spectrometer (LRS) slitless time-series mode \citep{kendrew_mid-infrared_2015}, with subarray \texttt{SLITLESSPRISM}, which covers a wavelength range of 5 -- 12\,$\mu$m. The visit was scheduled around the time expected for the secondary eclipse based on the transit ephemeris and assuming a circular orbit. The eclipse was detected at the expected time at 8$\sigma$ confidence in the white light curve, as described below. The visit began 2.6 hours before the eclipse, continued for the 46-minute eclipse duration, and concluded 1.2 hours after the end of the eclipse. A total of 3,594 integrations with 28 groups per integration (4.6\,s per integration) were obtained.

\section{Data Analysis} \label{sec:data_reduction}
To ensure the reproducibility of the results, we performed data analyses with both the \texttt{SPARTA} \citep[first described in][]{kempton_reflective_2023}\footnote{\url{https://github.com/ideasrule/sparta}} and \texttt{Eureka!} codes \citep{bell_eureka_2022}. These two JWST pipelines have shown remarkable agreement for MIRI and NIRCam data \citep{kempton_reflective_2023,xue_jwst_2024,zhang_gj_2024,powell_sulfur_2024,bell_w43_2024}. We present an overview of the analyses conducted by each pipeline in the following sections.

\subsection{\texttt{SPARTA}} \label{SPARTA_reduction}
\begin{figure*}[ht!]
\includegraphics[width = 18cm]{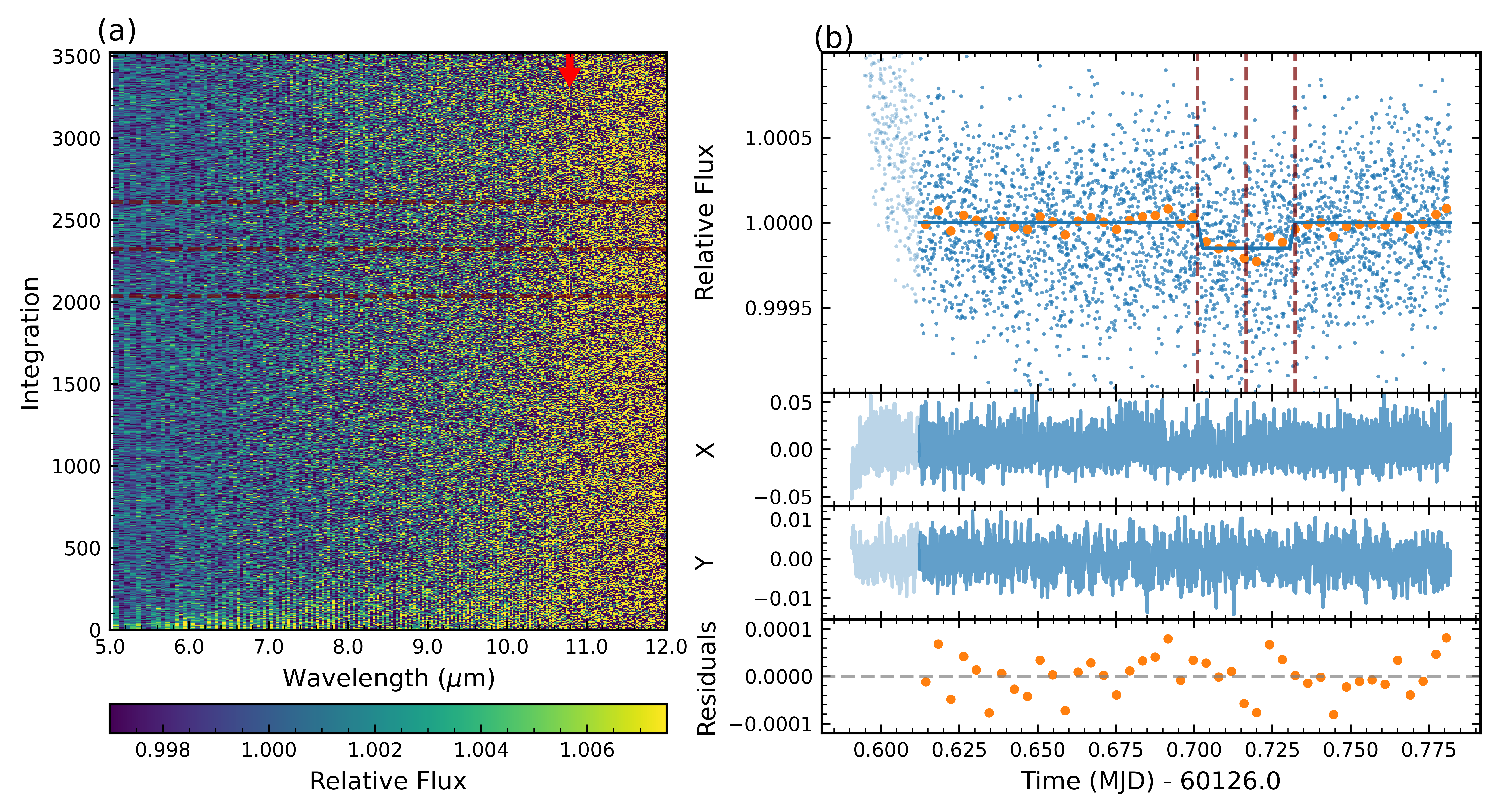} \\
\caption{Light curves produced by our nominal \texttt{SPARTA} reduction. (a) Raw 2D light curves in the wavelength-integration plane. Anomalies such as the odd-even effect and exponential downward ramps at the beginning $\sim$ 400 integrations of the observation ($\sim$ 30\,min) are found. The red arrow indicates the very bright strip (3$\times$ baseline flux) that we eventually removed from our nominal reduction. The start, middle, and end of the eclipse are shown with dashed dark red lines. (b) The systematics-corrected white light curve is shown in the top panel overplotted with the best \texttt{batman} eclipse model (blue line) and binned systematics-corrected fluxes (orange dots). The bottom panel shows the residuals between the best-fit total model($F_{sys}$ and $F_{batman}$) and binned fluxes. We excluded the first 400 integrations (30 minutes) due to the strong exponential ramp. The middle two panels show the drifts of the trace positions in the spatial and dispersion directions, which are decorrelated from the data with linear models. \label{fig:lightcurves}}
\end{figure*}

Starting from the \texttt{uncal.fits} files, \texttt{SPARTA} provides completely independent data reduction procedures, without using any codes from other existing pipelines. Our reduction is based on the most recent release of \texttt{SPARTA}, details of which can be found in \citet[Appendix A,][]{zhang_gj_2024}. Figure \ref{fig:lightcurves} shows the raw 2D light curves in the wavelength-integration plane and the systematics-corrected white light curve overplotted by the best eclipse model from our preferred reduction \texttt{SPARTA gr5} (see \S\ref{sec:interpretation}).

An anomalous downward offset of the last group in almost all the integrations was found in our data (more details can be found in \citealt{morrison_jwst_2023}), thus we excluded the last group from the up-the-ramp fitting. We also observed non-linear behaviors of the first $\sim$11 groups (appendix Figure~\ref{append_fig2}, \citealt{morrison_jwst_2023, dyrek_transiting_2024}). To test the impact of this non-linearity, we performed three independent analyses with \texttt{SPARTA}, removing zero, 5, and 11 groups from the up-the-ramp fitting, which we refer to as gr0, gr5, and gr11, respectively throughout the rest of this Letter. 

The center of the spectral trace is fixed at column 36 (determined by fitting it with a Gaussian curve) and a window with a half-width of 3 pixels is chosen for extraction. We adopted the result by simple extraction instead of optimal extraction because the latter introduced more scatter and inflated the median absolute deviation of the light curves. 

We identified a variety of systematics in the data. The two most evident systematics are the exponential downward ramp in the first $\sim$400 integrations and the odd-even effect (alternating bright/dark columns). Both of these have been seen in all the published MIRI/LRS transiting planet datasets \citep[e.g.,][]{bouwman_miri_2023, kempton_reflective_2023}. However, in our dataset, we do not see the so-called ``shadowed region effect'' spanning 10.5 -- 12\,$\mu m$ that has been discovered in some of the published datasets \citep{bell_w43_2024}. On the other hand, we found anomalous bright strips at 10.7861 and 10.8060\,$\mu$m (indicated by the red arrow in Figure~\ref{fig:lightcurves}(a)), the cause of which has not been determined. The pixel-level light curves at these two wavelengths are shown in Appendix Figure~\ref{append_fig4}. It is unlikely that they are caused by stellar activity or mirror tilt events \citep{schlawin_jwst_2023}. We ultimately removed the data at these two wavelengths. We performed a 4$\sigma$ rejection to the data based on fluxes and positions of the spectral traces, gathered data from 5 to 12\,$\mu$m, and binned the fluxes into 0.5\,$\mu$m bins to get spectroscopic light curves.

The systematics were detrended with:
\begin{equation} \label{eq1}
\begin{split}
    F_{sys} &= F_{star}(1+A e^{-t/\tau} + c_x x + c_y y \\
    &+ m(t-t_{mean})) 
\end{split}
\end{equation}

where $A$ is the exponential ramp amplitude, $\tau$ is the decay timescale; $x$ and $y$ are the positions of the trace in the dispersion and spatial directions, respectively (after rotating the raw images 90\degr);  and $c_x$, $c_y$ and $m$ are the linear terms to decorrelate drifts. The secondary eclipse was modeled with \texttt{batman} \citep{kreidberg_batman_2015}. We adopted the orbital period $P$=1.62892911 days, semi-major axis $a/R_{\star}$ = 15.2601, eccentricity $e$ = 0.0118, inclination $i$ = 88.16\degree, and argument of periapsis $\omega$ = -95.8\degree. These parameters were fixed for all the fitting procedures; see \S\ref{sec:stellar_parameters} and Table \ref{app_tab} for how we obtained them.  The eclipse time was determined by fitting the white light curve, and the best-fit value was used for the spectroscopic light curve fittings. We implemented the Markov chain Monte Carlo (MCMC) with \texttt{emcee} \citep{foreman-mackey_emcee_2012} to fit both the white light curve and the spectroscopic light curves in \texttt{SPARTA}.

\subsection{\texttt{Eureka!}}
Stages 1 and 2 in \texttt{Eureka!} are identical to those in the \texttt{jwst} pipeline \citep{bushouse_2022_7325378}. We followed the default \texttt{jump\_detection}$=4.0$ because changing it to $6.0$, $8.0$ or $12.0$ yielded the same white light secondary eclipse depth. In Stage 3, we defined the subarray region of interest as [80, 393] in the x direction and [10, 62] in the y direction. We applied the outlier rejection routine to the full frame with threshold [5,5] along the time axis, resulting in 0.01\% of the pixels flagged as bad. Optimal spectral extraction was performed with \texttt{spec\_hw}$=5$ pixels and \texttt{bg\_hw}$=10$ pixels. The spatial profiles were constructed by the median frame with an outlier rejection threshold $=10\sigma$. In Stage 4, we collected data from 5 to 12\,$\mu m$ and binned them into 14 spectroscopic light curves. Then we performed 4$\sigma$ outlier rejection with a box-car filter with \texttt{box\_width}  $= 15$. For the fitting, we utilized the dynamic nested sampling algorithms with the \texttt{dynesty} package\footnote{\url{https://dynesty.readthedocs.io/en/stable/}}. The light curves were fitted with a joint systematic model:

\begin{equation} \label{eq2}
\begin{split}
F_{sys} = & (c_0 + c_1 (t - t_{mean}))\times \\
& (1+r_0 e^{-(t_{min} - t_{min,0})/r_1}) \times \\
& (1 + c_y y) \times \\
& (1 + c_{\Delta \sigma_y} \Delta \sigma_y)
\end{split}
\end{equation}

where $c_0$ and $c_1$ are the linear decorrelation terms; $r_0$ and $r_1$ are the amplitude and the timescale of the exponential; $y$ is the drift or jitter in the spatial direction and $\Delta \sigma_y$ is the change in the PSF width in the spatial direction
Decorrelation in the dispersion direction was not included because we did not see strong trends with $x$ and $\Delta x$. We obtained the secondary eclipse model from \texttt{batman} with the $P$, $a/R_{\star}$, $i$, $e$, and $\omega$ fixed to the same values described above in the \texttt{SPARTA} section.

\section{Host Star and Planet Parameters} \label{sec:stellar_parameters}
Accurately assessing whether a planet has an atmosphere using thermal emission measurements hinges critically on knowledge of the planet and star parameters because the technique boils down to an energy balance argument \citep{koll_identifying_2019}. We ultimately need to know the amount of energy the planet receives, and we need to be able to convert a measured secondary eclipse depth into a planetary flux. The former depends on the stellar effective temperature (we adopt $T_s$ for this instead of the usual $T_{\mathrm{eff}}$ because we want to distinguish the stellar and planetary temperatures below) and the planet's semi-major axis in units of the stellar radius ($a/R_{\star}$). Converting the secondary eclipse depth into a planetary flux also requires the planet-to-star radius ratio ($R_p/R_{\star}$) and the stellar spectrum in the observed bandpass.


We therefore followed the procedure of \citet{Mahajan:2024} to use \exofasttwo \ \citep{Eastman:2019} to leverage the stellar density derived from JWST transit and secondary eclipse observations to obtain precise stellar and planetary parameters, as shown in Table \ref{tab:GJ1132.}. For our analysis of the GJ\,1132 system, we used the two primary transits from JWST published by \citet{may_double_2023}, one taken on February 25 2023 and the other on March 5 2023, as well the secondary eclipse presented in this paper (adopting the \texttt{SPARTA} \textsf{gr5} reduction, see next section). 
Note that for the GJ\,1132b primary transits, there were no processed white light curves available. The data were processed from the \texttt{Eureka!} Stage3 products available online\footnote{\url{https://zenodo.org/records/10002089}}. The data went through Stage4 using the same parameters as \citet{may_double_2023}. We also used the first sector of data (March 2019) from the Transiting Exoplanet Survey Satellite (TESS) to maximize the baseline between the TESS and JWST observations and therefore provide the best constraint on the period. While there were additional transits that could have constrained the period further, the addition of those data resulted in increased runtime for our models with minimal improvement in our measurement of the period. 

Our results are a dramatic improvement over the previous best analysis of the GJ\,1132 system by \citet{bonfils_radial_2018}. Notably, we measure the stellar radius 25\% more precise, the radius of planet b $2.7\times$ more precise, and the stellar temperature $2\times$ more precise. Also note that we do not assume a circular orbit. Rather, the timing and duration of the primary and secondary eclipses allow us to measure an eccentricity that is precisely near but not zero ($e=0.0118^{+0.047}_{-0.0099}$).

Unlike the analysis in \citet{Mahajan:2024}, we added the ability to model a ramp simultaneously to \exofasttwo, and included that for all JWST light curves. In addition, for this system, the primary and secondaries were not contiguous, which added the uncertainty in the period to our determination of the eccentricity from the eclipse timing. A major reason for the inclusion of the TESS light curves was to render this period uncertainty negligible. Note that \citet{dittmann_search_2017} found no evidence of significant transit timing variations (TTVs) for the GJ\,1132 system, so we assumed a linear ephemeris. 

Like \citet{Mahajan:2024}, we include radial velocity data \citep{bonfils_radial_2018} to measure the masses and eccentricity as precisely as possible. For this system, we simultaneously model all three planetary signals, including a radial velocity-only model for the non-transiting c planet and the d candidate, which may be stellar activity \citep{bonfils_radial_2018}.

Typically, EXOFASTv2 runs until the Gelman-Rubin statistic is less than 1.01 and the number of independent draws is greater than 1000 for each parameter. In this fit, however, a handful of parameters did not pass this strict convergence criteria. In particular, the masses of the two planets and the ramp parameters for the eclipse detrending had Gelman-Rubin statistics of ranging from 1.05-1.09 and independent draws ranging from 800-1200. We expect the impact on the inferred median values and uncertainties to be negligible \citep[][Eq 35]{Eastman:2019}, and these parameters are not strongly covariant with the parameters used in subsequent analyses.

The posterior of the MCMC sampling of this global model is subsequently used to  account for the correlation between stellar and planet parameters when deriving dayside brightness temperature of the planet, as is discussed in \S\ref{sec:albedo}.

\section{Interpretation} 
\label{sec:interpretation}
\begin{figure}[ht!]
\includegraphics[width = 8.5cm]{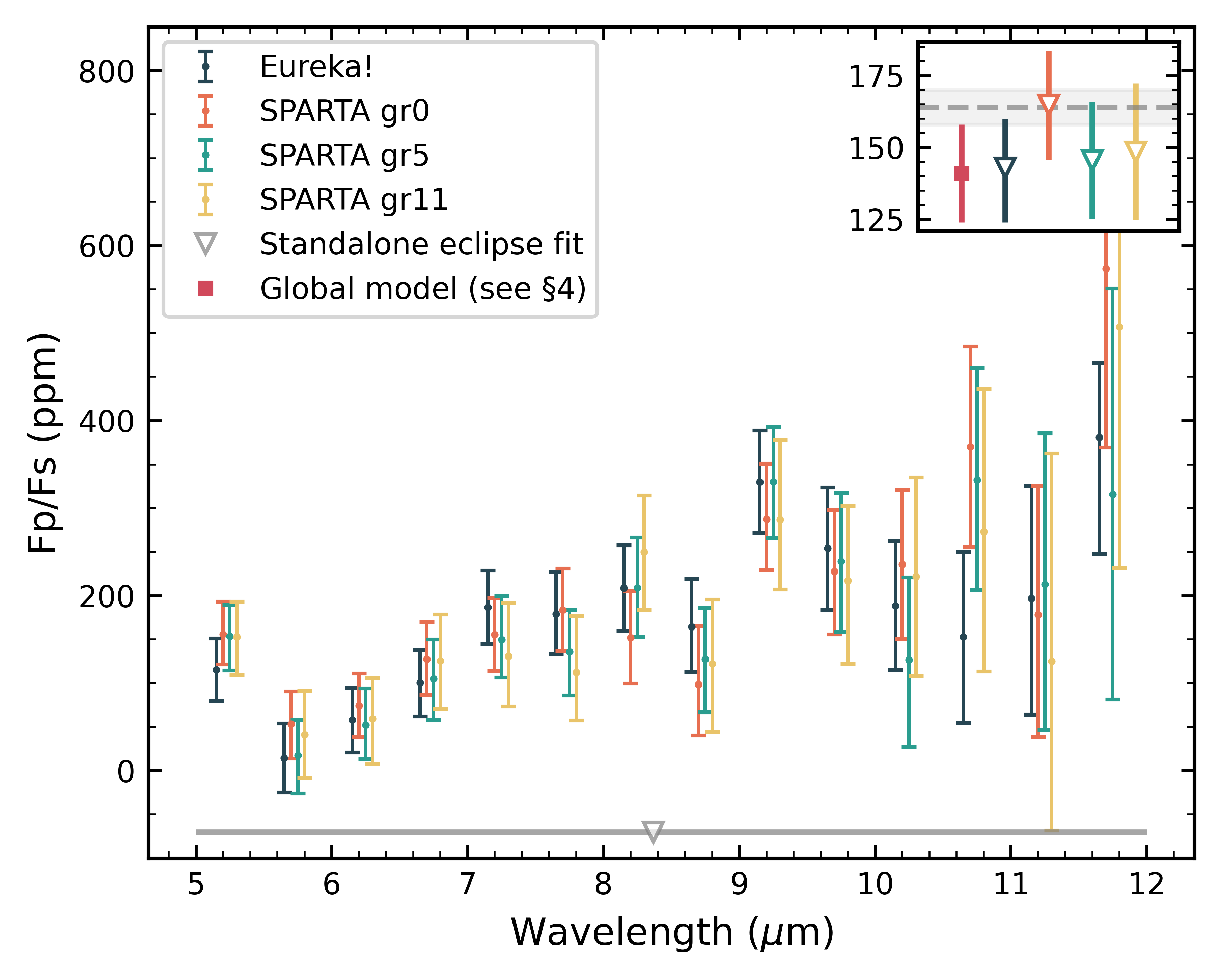} \\
\caption{Emission spectra (main panel) and white-light eclipse depths (inset panel) from the different reductions. The results from \texttt{Eureka!} are shown in dark blue, and the results from \texttt{SPARTA} with different up-the-ramp fitting strategies are presented in pink (using all groups), green (with the first 5 groups removed) and yellow (with the first 11 groups removed). The wavelengths are shifted for visualization purposes. The grey dashed line in the inset panel shows eclipse depth corresponding to the maximum dayside temperature of the planet with the shaded area indicating $\pm 1\sigma$. The white-light bandpass is shown with the gray triangle in the main plot. We plot the eclipse depth measured by the global model described in \S \ref{sec:stellar_parameters} with the red square. 
\label{different_reductions}}
\end{figure}

Figure~\ref{different_reductions} shows a comparison of the results from the different reductions and modeling approaches. We found good agreement in the spectra and white light eclipse depths between all these analyses. The most discrepant result is the white-light eclipse depth from the \texttt{SPARTA} \texttt{gr0} reduction, which is 10\% higher than the average of the other four, but still consistent with the others within the 1$\sigma$ range. The average error of the eclipse depths of the \texttt{SPARTA} reductions with more groups trimmed at the beginning is larger due to the loss of information, but we consider the results of these reductions more robust. Notably, the lower bound of $Fp/Fs$ in the \texttt{gr11} reduction at 11.25\,$\mu m$ is negative, which is unphysical. Due to the significant non-linear trend we observed in the first $\sim$5 groups (appendix Figure~5), and the error inflation on the spectrum in the \texttt{gr11} reduction, we decided to use the \texttt{SPARTA} \textsf{gr5} reduction as our preferred version. Combining with the \S \ref{sec:stellar_parameters} analysis, we report the white-light eclipse depth to be 140\,$\pm$\,17\,ppm. We perform two kinds of modeling below to interpret these results.

\subsection{Inferred Albedo and Heat Recirculation Efficiency}
\label{sec:albedo}
In this section, we derive the dayside temperature of GJ\,1132b from the measured white light eclipse depth assuming the planet is a perfect blackbody with a smooth surface. That is, we determine the planet's brightness temperature over the broad 5 -- 12 $\mu$m MIRI/LRS band. We then compare this temperature to the maximum possible temperature and use that comparison to put joint constraints on the Bond albedo (but see caveats at the end of this section) and heat recirculation efficiency.

Deriving the planet's brightness temperature involves inverting the $F_p/F_s$ determined from the secondary eclipse. This is usually done by fitting the measured $F_p/F_s$ with a model (e.g., using non-linear least squares). However, we aimed to account for the correlated uncertainties in the eclipse depth and the star and planet parameters. We also aimed to account for the correlated uncertainties between the measured planet temperature and the maximum possible temperature. We took advantage of the posteriors from the MCMC in \S\ref{sec:stellar_parameters} to do this. However, it required a different approach than the usual one because there are 557,999 steps in the Markov chains and it would be prohibitively expensive to invert each one by fitting. We therefore follow the procedure below to derive the planet's brightness temperature.

The eclipse depth can be modeled by dividing planetary spectra by stellar spectra weighted by instrument throughput in units of photon numbers:
\begin{equation}\label{temp_eq}
\begin{split}
\frac{F_p}{F_s} 
& = \left(\frac{R_p}{R_s}\right)^2 \cdot \frac{\int \frac{\pi \cdot B_p(T_p, \lambda)}{hc/\lambda} \cdot W_{\lambda}\, d\lambda}{\int \frac{M_{s}(T_s,\,\mathrm{log}\,g,\,[M/H],\,\lambda)}{hc/\lambda}\cdot W_{\lambda} \, d\lambda}
\end{split}
\end{equation}
where $W_\lambda$ is the throughput of MIRI/LRS\footnote{available on Zenodo \dataset[10.5281/zenodo.13244543]{https://doi.org/10.5281/zenodo.13244543}}, $B_p(T_p, \lambda)$ is the planetary blackbody intensity and $M_{s}(T_s,\,\log g,\,[M/H]\footnote{[M/H] = $\log \bigl[ \frac{n(Metal)}{n(H)} \bigr] - \log \bigl[ \frac{n(Metal)}{n(H)} \bigr]_{\odot}$},   \,\lambda)$ is the stellar spectrum. For stellar spectra, we used both models interpolated from the PHOENIX grid \citep{phoenix} using the python package \texttt{pysynphot}\footnote{\url{https://pysynphot.readthedocs.io/en/latest/}} \citep{pysynphot} and models from the SPHINX grid \citep[][updated on May 30, 2024]{iyer_sphinx_2023} interpolated by our own codes. The two model grids gave nearly identical results; we adopted the PHOENIX results below.

To take into account instrumental broadening, we convolved the model stellar spectra using a Gaussian kernel with FWHM identical to MIRI/LRS's resolution\footnote{R$\sim$100 at $7.5 \mu m$}. We converted blackbody intensity to blackbody flux by multiplying $B_p(T_p, \lambda)$ with $\pi$. Before multiplying by the instrument throughput, the spectra were divided by $hc/\lambda$ to convert from energy to photon flux. The integrals in Equation~\ref{temp_eq} are calculated from 5 to 12 $\mu m$.

We obtain $\frac{F_p}{F_s}$, $\frac{R_p}{R_s}$, [M/H], $\log g$, and $T_s$ from the MCMC chain sampled when fitting the global model to the white light curve (see \S \ref{sec:stellar_parameters}). For each sample in the chain, we are able to invert Equation \ref{temp_eq} to determine the photon flux emitted by the planet at its dayside.

Finally, to convert the photon flux emitted by the planet to the dayside brightness temperature without expensive fitting processes, we prepared a photon\,flux-T$_p$ grid calculated with $F_{photon} = \int \frac{\pi \cdot B_p(T_p, \lambda) \cdot W_{\lambda}}{hc/\lambda}\, d\lambda$ for $T_p$ ranging from 300 to 1,000\,K. We then interpolate $T_{p,\mathrm{dayside}}$ from this grid. This thus gives us posterior samples for $T_p$ that are correlated with the star and planet parameters. From this analysis we find that the brightness temperature of the planet's dayside is 709\,$\pm$\,31\,K.

\begin{figure*}[ht!]
\includegraphics[width = 18cm]{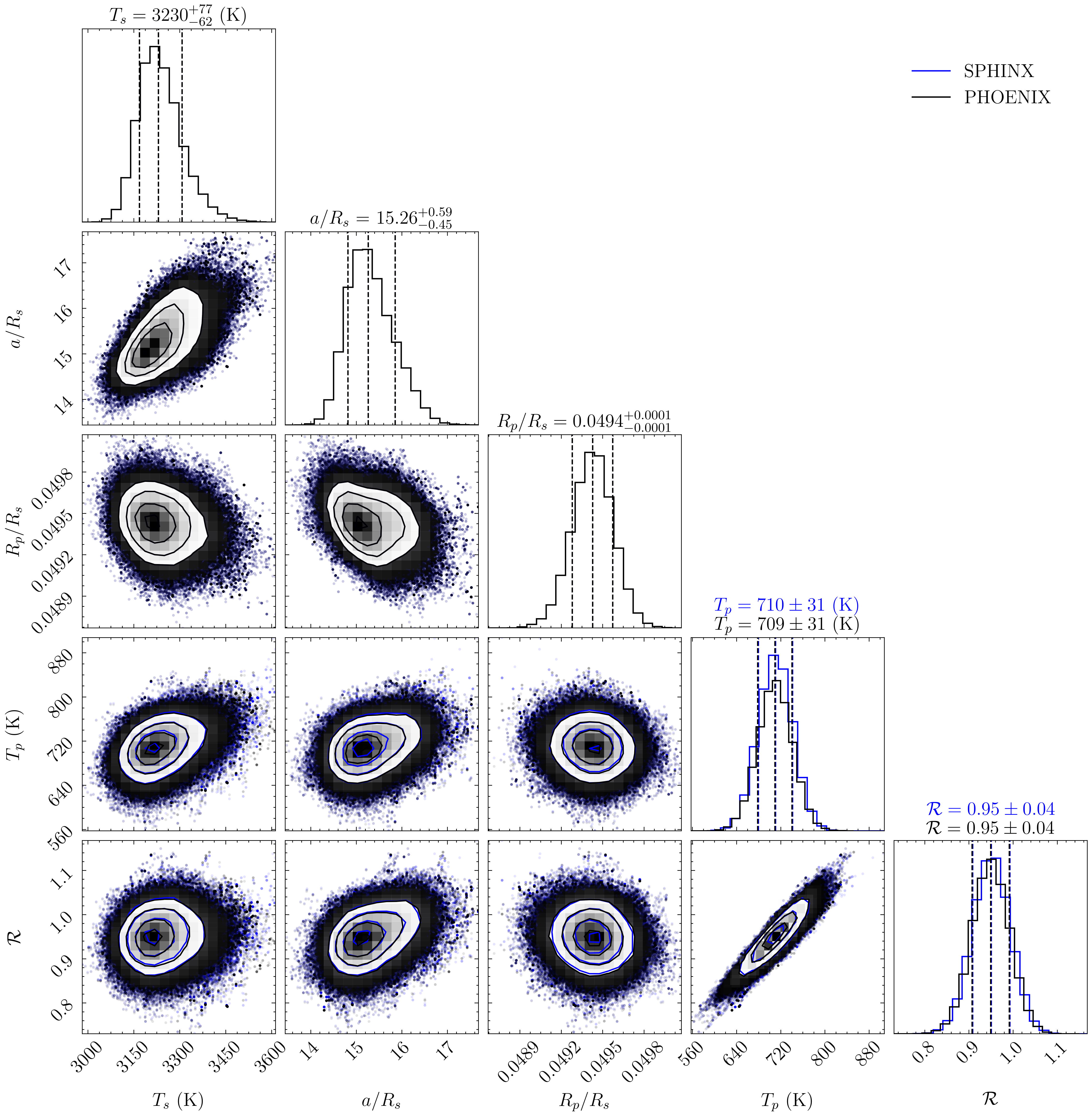} \\
\caption{Corner plot of the planet temperature and temperature factor along with key star/planet parameters. The posteriors for $T_s$, $a/R_{\star}$, and $R_p/R_{\star}$ come from the global modeling described in \S\ref{sec:stellar_parameters}. The posteriors for $T_p$ and $\mathcal{R}$ come from the modeling described in \S\ref{sec:albedo}. Black are the results using PHOENIX stellar models, and blue are the results using SPHINX models.
\label{fig:Tp_corner}}
\end{figure*}

To compare the determined $T_{p,\mathrm{dayside}}$ with the maximum possible dayside brightness temperature $T_{max}$, we define the temperature scaling factor $\mathcal{R}$ by

\begin{equation} \label{tempfactor_eq}
\begin{split}
T_{p,dayside} & = T_{max} \cdot \mathcal{R} \\
& = \left( \frac{2}{3} \right)^{\frac{1}{4}} \cdot \frac{T_s}{\sqrt{a/R_s}}\cdot \mathcal{R} 
\end{split}
\end{equation}

Following \citet{cowan_statistics_2011}, $\mathcal{R}$ is defined as 
\begin{equation}\label{albedo_e}
    \textstyle
    \mathcal{R} =  \left( \frac{2}{3} \right)^{-\frac{1}{4}} \cdot (1-A_B)^{\frac{1}{4}}\cdot(\frac{2}{3}-\frac{5}{12}\varepsilon)^{\frac{1}{4}} 
\end{equation}
where $A_B$ is the Bond albedo and $\varepsilon$ is the heat recirculation efficiency. Note the $\left( \frac{2}{3} \right)^{\frac{1}{4}}$ factors in Equations \ref{tempfactor_eq} and \ref{albedo_e} are for normalization purpose, so when $A_B = 0$ and $\epsilon = 0$ (zero reflection and no heat redistribution), $\mathcal{R} = 1$, representing the maximum possible dayside temperature, $T_{max} = 746^{+14}_{-11}$\,K.

We calculate $\mathcal{R}$ for each sample in the MCMC chain derived in \S~\ref{sec:stellar_parameters}. We present the corner plots of relevant stellar parameters, planetary dayside temperature and associated $\mathcal{R}$ in Figure~\ref{fig:Tp_corner}. From the measured white light eclipse depth, we derive a temperature scaling factor of $\mathcal{R} = 0.95 \pm 0.04$ 

\begin{figure}[ht]
\includegraphics[width = 9cm]{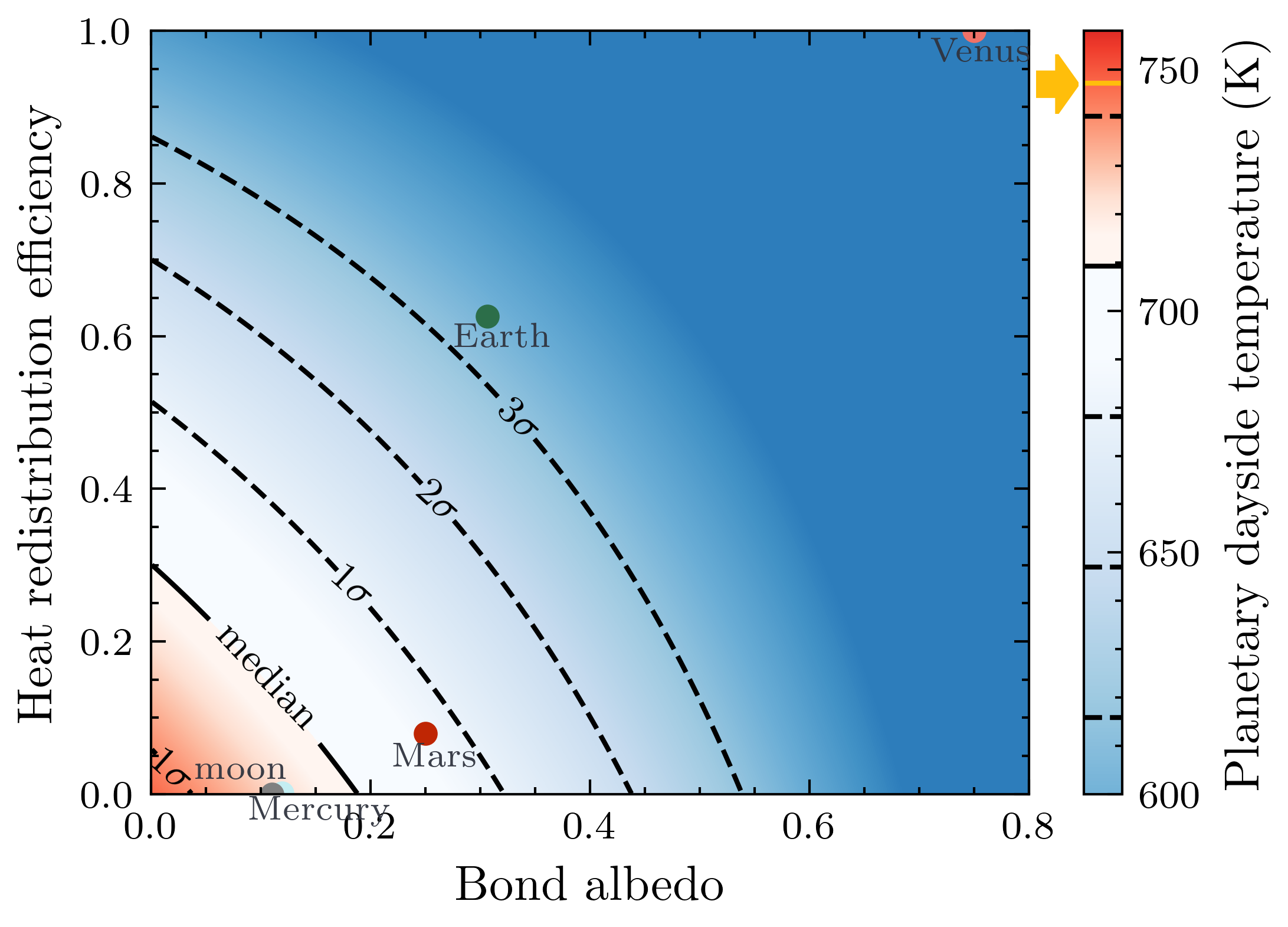} \\
\caption{Planetary dayside temperature as a function of Bond albedo $A_B$ and heat recirculation efficiency $\varepsilon$. The median and $\pm 1\sigma$ contours show the $709\pm31$\,K dayside temperature derived from the white light curve. The yellow arrow indicates the maximum dayside temperature (746$^{+14}_{-11}$\,K) assuming no reflection and no heat recirculation. We adopt $A_B$ = 0.119, 0.75, 0.306, 0.250, 0.11 for Mercury, Venus, Earth, Mars, and the Moon. Since the solar system terrestrial planets are not tidally locked, we calculated their heat redistribution efficiency using their polar and equatorial temperatures. Following equation 4 \& 5 in \citet{cowan_statistics_2011}, we derived $\epsilon = \frac{8/3}{((T_{\mathrm{equator}}/T_{\mathrm{pole}})^4 +5/3)}$. \label{fig:map}}
\end{figure}

Combining equations \ref{tempfactor_eq} \& \ref{albedo_e}, we can get $T_{p,\mathrm{dayside}}$ as a function of $A_B$ and $\varepsilon$. To explore how these two factors affect $T_{p,\mathrm{dayside}}$, we adopted the median values of $T_s$ and $a/R_s$ from the MCMC (see Table \ref{tab:GJ1132.}), and varied $A_B$ and $\varepsilon$ from 0 to 1 to calculate $T_{p,\mathrm{dayside}}$ as a function of $A_B$ and $\varepsilon$. Afterwards, we plotted the map of $T_p(A_B, \varepsilon)$ in Figure~\ref{fig:map} with the uncertainties of $T_{p, \mathrm{dayside}}$ shown with contours.

Both the derived $\mathcal{R}$ and Figure \ref{fig:map} indicate that our inferred dayside temperature for GJ\,1132b is only $\sim1\sigma$ lower than the maximum possible temperature (746$^{+14}_{-11}$\,K). This suggests that the planet has little to no day-to-night heat redistribution and that the surface is relatively dark. Thus the data are consistent with the presence of no atmosphere on the planet. This is reinforced by comparing the joint constraints on the Bond albedo and heat redistribution factor with the values for terrestrial objects in the solar system in Figure~\ref{fig:map}. The results for GJ\,1132b are consistent with the parameters of the airless and nearly-airless bodies (i.e., Mercury, Mars, and Earth's Moon), and they are inconsistent with the parameters of the planets that have significant atmospheres (i.e., Earth and Venus).

To place quantitative limits on the thickness of a possible atmosphere, we followed the approach in \citet{zhang_gj_2024}, who used the scaling relation for the heat redistribution parameter in \citet{koll_scaling_2022}. We adopted an upper limit on the heat redistribution efficiency of 0.52, which is the 1$\sigma$ upper limit from our constraints when fixing $A_B$=0. We find an upper limit of 0.7\,bar for a pure H$_2$O atmosphere, and 2.4\,bar for a pure CO$_2$ atmosphere. Thus, we can rule out very thick, Venus-like atmospheres solely on energy transport arguments alone. Earth-like, 1\,bar atmospheres are marginally consistent with our constraints on the heat redistribution. However, such atmospheres would also likely have a non-zero Bond albedo, as illustrated in Figure~\ref{fig:map}.

A caveat to this analysis is that we do not consider thermal beaming, which may result from (e.g.) surface roughness effects \citep{spencer_1990}, and which can increase the observed low-phase-angle brightness temperature of atmosphere-free bodies above the maximum stated above \citep{emery_1988}. On the other hand, even a thin atmosphere would likely negate thermal beaming, thus making our atmospheric thickness constraints still valid.

Another important caveat to this analysis is that we are actually only constraining the effective albedo over the MIRI/LRS bandpass rather than the true Bond albedo. A blackbody with the dayside temperature we infer for GJ\,1132b (709\,K) emits 48\% of its total energy between 5 and 12\,$\mu m$. Therefore, we are likely capturing a large fraction of the energy emitted by the planet, and this is a key aspect of estimating the Bond albedo \citep[see discussion in][]{kempton_reflective_2023}. On the other hand, even bare surfaces will have wavelength-dependent albedos, and the most likely minerals on GJ\,1132b have lower-than-average albedos in the MIRI/LRS bandpass \citep{hu2012}. As shown by \citet{mansfield_identifying_2019}, using MIRI/LRS can lead to underestimating the Bond albedo of GJ\,1132b by 0.1 -- 0.2, while \citet{Whittaker_lhs3844_2022} suggested this underestimation is less serious. However, space weathering of close-in planets like GJ\,1132b will serve to darken surfaces and make them more blackbody like \citep[][and references therein]{zieba_no_2023,lyu2024}, thus making our assumptions more valid. With these caveats, if we assume GJ\,1132b doesn't have an atmosphere, then our constraints yield $A_B=0.19^{+0.12}_{-0.15}$.




\subsection{Forward Modeling}

\begin{figure*}
    \includegraphics[width=\textwidth]{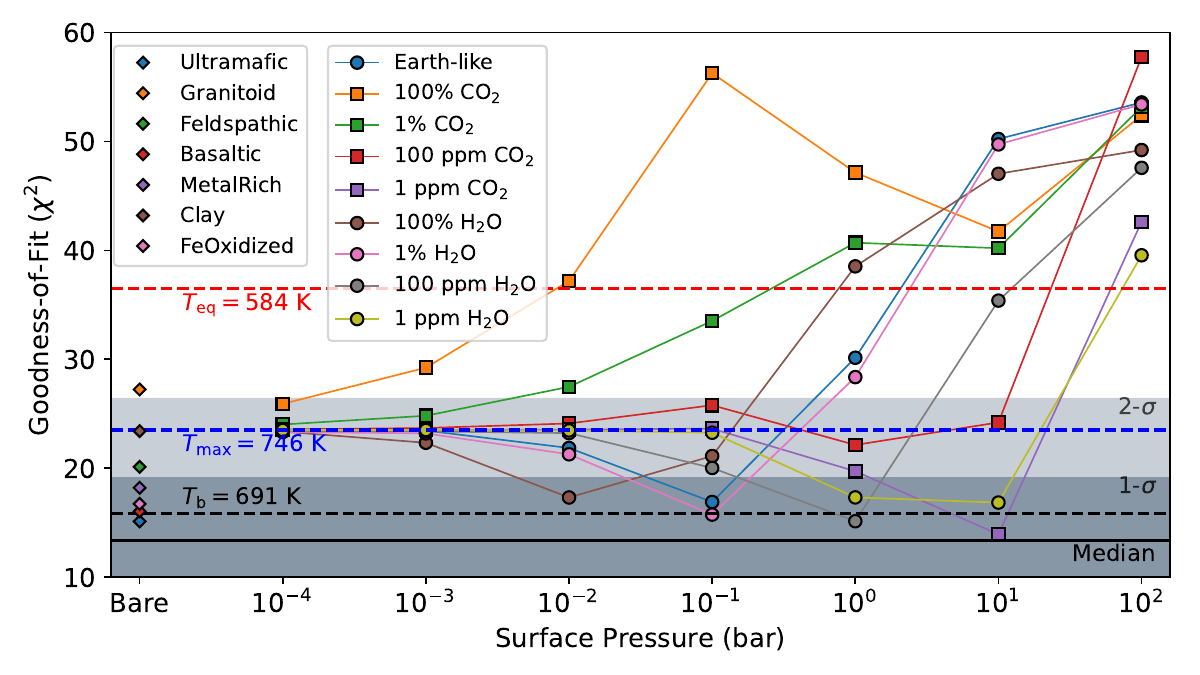}
    \caption{Goodness-of-fit ($\chi^2$) of the eclipse depths of the suite of atmospheric models, plotted as functions of the modeled surface pressure. The best-fit blackbody (691\,K) to the emission spectrum is shown with the black dashed line. The bare surface models are shown in the leftmost column with diamonds.  The median and 1-$\sigma$ and 2-$\sigma$ equivalent quantiles of the $\chi^2$ distribution ($k$=14, $k$ is the degree of freedom) are shown as the solid black line, the dark gray band, and the light gray band, respectively.  The atmosphere models with compositions of CO$_2$ filled with O$_2$ are shown with squares and H$_2$O filled with O$_2$ are shown with circles, where colors indicate the percentage of trace gasses. The atmosphere models generally move, from the thinnest ($10^{-4}$ bar) to the thickest ($10^2$ bar) surface pressures, from the $\chi^2$ for the maximally hot temperature (blue dashed line) to the $\chi^2$ for the equilibrium temperature (red dashed line), with deviating inflection points arising due to spectral features in the LRS bandpass and changes in the temperature-pressure profile.  Atmospheric models that produce $\chi^2$ values comparable to the best-fit blackbody are plotted in Figure~\ref{fig:tb_spectra}.}
    \label{fig:psurf_chisq}
\end{figure*}

\begin{figure*}
    \includegraphics[width=\textwidth]{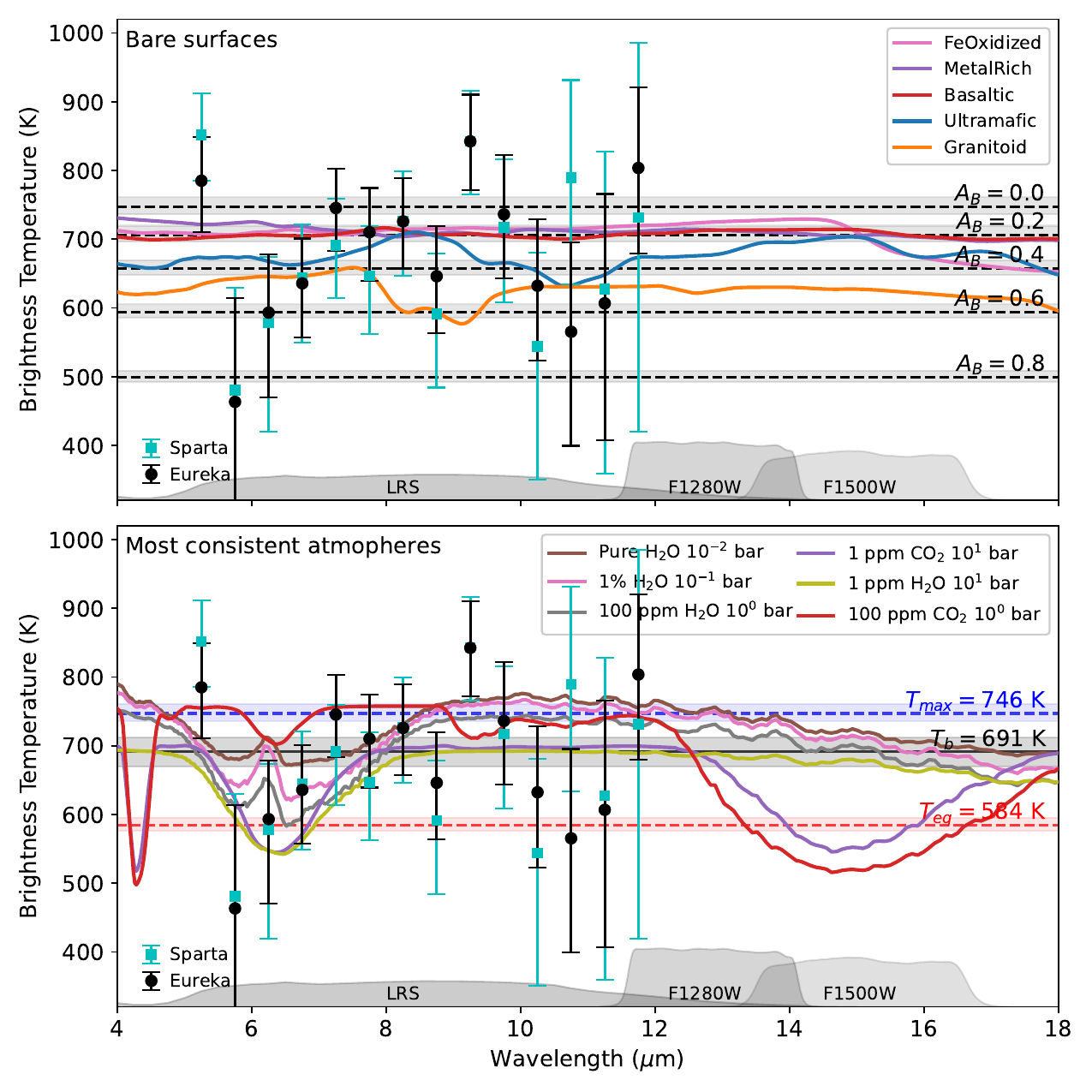}
    \caption{Observed brightness temperature spectrum from \texttt{SPARTA} (cyan) and \texttt{Eureka!} (black) plotted against models for bare surfaces (top panel) and atmospheres (bottom) that fit the data.  The expected brightness temperatures and their uncertainties propagated from the system parameters are plotted as horizontal lines for varying Bond albedos and redistribution factors.  The bandpass functions for the LRS and the 12.8 and 15\,$\mu$m MIRI photometry filters are also shown for illustration. As can be seen from 1 ppm and 10 ppm CO$_2$ models, the presence of CO$_2$ can be further constrained using 15\,$\mu$m MIRI photometry.}
\label{fig:tb_spectra}
\end{figure*}

Following the methods in \citet{Whittaker_lhs3844_2022} and \citet{Ih_T1b_2023}, we used \texttt{HELIOS} \citep{helios1, helios3, helios2} to compute forward models to compare to the dayside thermal emission spectrum of GJ\,1132b.  Complementing the white light analysis, this allows for incorporating the spectral information to test which atmosphere models of varying surface pressure and composition are plausible.  We describe the model set up and present which model thermal emission spectra are consistent with the spectra here. 

We adopted the median system parameters derived in \S\ref{sec:stellar_parameters}, and we generated the stellar spectrum using the SPHINX model grid \citep{iyer_sphinx_2023}.  We calculated eclipse spectra for atmosphere models of varying surface pressure from 10$^{-4}$ to 10$^2$ bar in 1-dex intervals (assuming a blackbody surface under the atmosphere) and compositions of 1 ppm, 100 ppm, 1\%, and 100\% CO$_2$ or H$_2$O, with the rest of the composition backfilled with O$_2$.  We note that the choice of backfilling gas between O$_2$ and N$_2$ does not affect our results.  We also generated eclipse spectra for bare surface models using the surface albedo spectra from \citet{hu2012}.


We determined the brightness temperature in the observed eclipse depth spectrum by using the \texttt{emcee} package to find the temperature that minimizes the least-square fit between the data and a blackbody model.  From this we obtain $T_{\rm{b}} = 691 ^{+22}_{-21}$ K. This is smaller than the temperature derived from the white light curve because of the difference in how the channels are weighted. In the white light curve, the channels are summed and are thus weighted by photon counts. In the spectrum, the average is computed by weighting by the errors on the different channels. The errors in the spectroscopic channels are relatively larger compared to the stellar photon counts at longer wavelengths because of the increased background in the raw data and the higher levels of red noise in the light curves. Thus the longer wavelength channels, which suggest a higher brightness temperature, are de-weighted in the spectroscopic analysis here. Furthermore, the error bar on the temperature derived from the spectra is smaller than from the white light curve because it doesn't account for uncertainties in the star and planet parameters. Importantly, this is better than 1$\sigma$ consistent with the T$_\mathrm{b}$ from our broadband measurement(\S \ref{sec:albedo}), further validating our analytical approach.

From the modeled eclipse depth spectra, we computed the goodness-of-fit ($\chi^2$) to the data for each model, which we show in Figure~\ref{fig:psurf_chisq}.  As a function of the surface pressure, the $\chi^2$ values of atmosphere models are controlled by three effects: (1) cooling by redistribution from day to nightside; (2) spectral features from molecular absorption; (3) greenhouse warming. This leads to non-monotonic behavior in $\chi^2$ and potentially multiple inflection points.

We find that a broad range of atmospheres are consistent with the data.  The atmospheres that are readily ruled out at $>3\sigma$ are either too thick (all of the 100\,bar atmospheres and most of the 10\,bar atmospheres) or have too much molecular absorption (pure CO$_2$ atmospheres above 0.006\,bar or pure H$_2$O atmospheres above 0.16\,bar).  Compared to the white-light analysis, this sets the upper limit for the pure CO$_2$ or H$_2$O atmospheres at more tenuous surface pressures, as now the shape of the spectrum is taken into account.  As such, the spectrum now rules out even a Mars-like thin atmosphere, if composed entirely of CO$_2$. 

The thinnest atmospheres modelled have $\chi^2$ values very close to that of the blackbody at $T_{\rm{max}}$ (blue dashed line), while the thickest surface pressure modelled have worse $\chi^2$ values than that of the blackbody at $T_{\rm{eq}}$ (red dashed line).  At intermediate surface pressures of $10^{-1}-10^1$ bar, the atmosphere models show comparable $\chi^2$ values to the best-fitting blackbody (black dashed line), but do not show an appreciable improvement in goodness-of-fit because the blackbody already explains the data well.

For these models, we show the brightness temperature spectra in Figure~\ref{fig:tb_spectra}.  The brightness temperature of the data is obtained by first multiplying the binned model stellar spectrum by the observed $F_{\rm{p}}/F_{\rm{s}}$ spectrum and propagating the system parameter uncertainties to the errors.  The few points in the data that have low brightness temperature around 6\,$\mu$m align with either CO$_2$ or H$_2$O features, but the errors are large enough to preclude confidently inferring that an absorption feature is present. We also note that the transmission spectra of modeled atmospheres are consistent with the featureless transmission spectrum in \citet{may_double_2023}.

Similarly, we find that a broad range of bare surface models are consistent with the data.  We show the goodness-of-fit for each model in Figure~\ref{fig:psurf_chisq} and the brightness temperature spectrum in Figure~\ref{fig:tb_spectra}.  For these models, the $\chi^2$ values are controlled primarily by their Bond albedo in the shortwave affecting the dayside temperature and secondarily by the spectral emissivity in the MIRI bandpass.  As with the atmosphere models, the errors are large enough to preclude inferring a spectral feature.


\section{Discussion} \label{sec:discussion}
In this paper, we reported a secondary eclipse observation of the super-Earth GJ\,1132b with JWST MIRI/LRS, yielding a measurement of the white-light dayside brightness temperature and dayside emission spectrum. We also refined the star and planet parameters for this benchmark system. Given energy balance, the measured secondary eclipse ($140 \pm 17$\,ppm) is very close to the maximum possible depth ($164\pm~6$\,ppm). The dayside emission spectrum exhibits no significant spectral features and is consistent with a blackbody that has T$_\mathrm{b}$\,=\,691$^{+22}_{-21}$\,K. 

We compared the dayside emission spectrum with a wide range of atmospheric composition models. We found atmospheres with thickness $P\,>\,$1\,bar and containing at least 1\% H$_2$O are ruled out, and atmospheres of any modeled thickness (10$^{-4}$ bar -- 10 $^2$ bar), containing at least 1\% CO$_2$ are ruled out. We also found very thick, Venus-like atmospheres ($P\,\sim $ 10$^{2}$\,bar), even without significant infrared absorbers, are ruled out. A few points around 6 $\mu$m could suggest CO$_2$ or H$_2$O features, but the errors are too large to infer conclusive results. Due to their high mean molecular weight, the transmission spectra produced by all modeled atmospheres show no significant features, which is consistent with \citet{may_double_2023}. From the bare surface models, we found a wide range of possible surface compositions are consistent with the data. Thus we conclude that, given the preponderance of the evidence, the planet likely does not have a significant atmosphere.

\begin{figure}[t]
\includegraphics[width = 9cm]{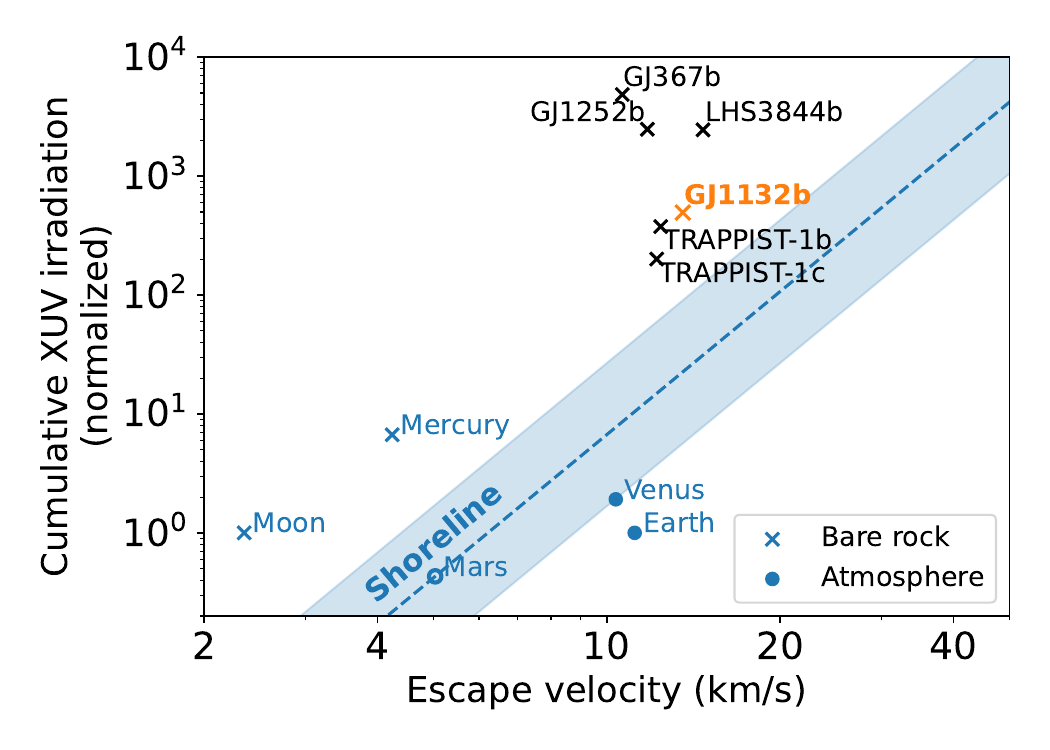} \\
\caption{GJ\,1132b in the context of the Cosmic Shoreline \citep{zahnle_cosmic_2017}. In the solar system, the dashed blue line separates bodies that have atmospheres
from those that don’t. Blue shading illustrates uncertainty on the y-axis. For exoplanets, only rocky planets around M-dwarfs with thermal emission measurements} are shown. \label{fig:cosmic_shoreline}
\end{figure}

These results for GJ\,1132b provide an additional exoplanet data point with which to test the ``Cosmic Shoreline'' theory \citep{zahnle_cosmic_2017}. This theory suggests a planet's ability to maintain an atmosphere is based on its predicted escape velocity (v$_{esc}$) and its cumulative extreme UV (XUV) irradiation. Figure~\ref{fig:cosmic_shoreline} shows the Cosmic Shoreline for a number of Solar System bodies and rocky M-dwarf planets. The normalized cumulative XUV irradiation (normalized to Earth) is estimated using an analytic scaling \citep[Eq.\ 27 in][]{zahnle_cosmic_2017}, and the dashed line is a power law fitted to match Mars. The actual XUV irradiation history of particular exoplanets is unknown, so the blue region shows an illustrative factor of 4 uncertainty. Higher escape velocities and less insolation are more favorable for atmospheric survival. Our target, however, is placed more than 1$\sigma$ above the shoreline (Figure \ref{fig:cosmic_shoreline}). Further away from the shoreline are the three M-dwarf planets likely to have no atmospheres based on Spitzer and JWST observations \citep[LHS\,3844b, GJ\,1252b, and GJ\,367b:][]{kreidberg_absence_2019,crossfield_gj_2022,lyu_super-earth_2024,zhang_gj_2024}. Recent JWST MIRI observations of TRAPPIST-1 b and c \citep{greene_thermal_2023, zieba_no_2023} also suggest that neither of them have thick atmospheres. The relative position of these planets on the Cosmic Shoreline reinforces our conclusion that GJ\,1132b is most likely to be a bare rock with no or a thin atmosphere, consistent with what we see in Figure~\ref{fig:map}.

The question of whether M dwarf rocky planets can host atmospheres remains unresolved. Despite their higher risk of losing the atmospheres than for rocky planets around Sun-like stars, as we discussed in \S\ref{sec:intro}, there are still reasons to be optimistic. The planets might be able to accumulate a water-rich envelope beyond the snowline during formation, then migrate inward, or have an H$_2$ layer that could shield water from loss \citep[e.g.][]{ribas_habitability_2016,barnes_habitability_2018,kite_schaefer2021}, but see also \cite{kite_exoplanet_2020}. Other hypotheses include having magnetic fields that would guard against some loss mechanisms \citep[e.g.][]{segura_effect_2010,vidotto_effects_2013}, starting rich in carbon derived from refractory organics \citep{li_carbon_2021}, renewing their secondary atmospheres from outgassing after an initial loss phase \citep{kite_exoplanet_2020}, retaining atmospheres against efficient loss due to atomic line radiative cooling \citep{nakayama_survival_2022}, or being resupplied with volatiles from an external source, such as through cometary bombardment. Although our GJ\,1132b observation indicates GJ\,1132b likely does not have any atmosphere, our result further constrains the possible location of the Cosmic Shoreline. We hope future observations, such as the proposed 500~hrs JWST DDT program \citep{redfield24}, will further refine it.

\section{Data Availability}
\label{data_availability}
The data presented in this paper were obtained from the Mikulski Archive for Space Telescopes (MAST) at the Space Telescope Science Institute. The specific observations analyzed can be accessed via \dataset[DOI: 10.17909/br8w-4288]{https://doi.org/10.17909/br8w-4288}. The data that were used to create all of the figures will be freely available on Zenodo \citep{xue_2024_13244543}\dataset[DOI: 10.5281/zenodo.13244543]{https://doi.org/10.5281/zenodo.13244543}. All additional data is available upon request.

\facilities{JWST(MIRI)}


\software{\texttt{Eureka!} \citep{bell_eureka_2022},  
          \texttt{SPARTA} \citep{kempton_reflective_2023}, 
          \texttt{Astropy} \citep{astropy:2013,astropy:2018,astropy:2022}, 
          \texttt{emcee} \citep{foreman-mackey_emcee_2012}, 
          \texttt{dynesty} \citep{speagle_dynesty_2019}, 
          \texttt{batman} \citep{kreidberg_batman_2015},
          \texttt{HELIOS} \citep{helios1,helios2,helios3,Whittaker_lhs3844_2022}, 
          \texttt{Matplotlib} \citep{Hunter:2007}, 
          \texttt{Numpy} \citep{harris2020array}, 
          \texttt{Scipy} \citep{2020SciPy-NMeth},
          texttt{EXOFASTv2} \citep{Eastman:2019}
          }

\section{Acknowledgements}

This work is based on observations made with the NASA/ESA/CSA JWST. The data were obtained from the Mikulski Archive for Space Telescopes at the Space Telescope Science Institute (STScI), which is operated by the Association of Universities for Research in Astronomy, Inc., under NASA contract NAS 5-03127 for JWST. The observations are associated with program GTO 1274.  MZ is grateful for support from the 51 Pegasi b Fellowship, funded by the Heising-Simons Foundation. MWM acknowledges support through the NASA Hubble Fellowship grant HST-HF2-51485.001-A awarded by STScI. JL was supported by grant NNX17AL71A from NASA Goddard Spaceflight Center. J. I. and E. M.-R. K. acknowledge support from the AEThER Matter-to-Life program, funded by the Alfred P. Sloan Foundation under grant G202114194.

\newpage
\appendix
\renewcommand{\thefigure}{A\arabic{figure}}
\renewcommand{\thetable}{A\arabic{table}}
\section{Appendices}

\begin{figure}[ht!]
\includegraphics[width = 9cm]{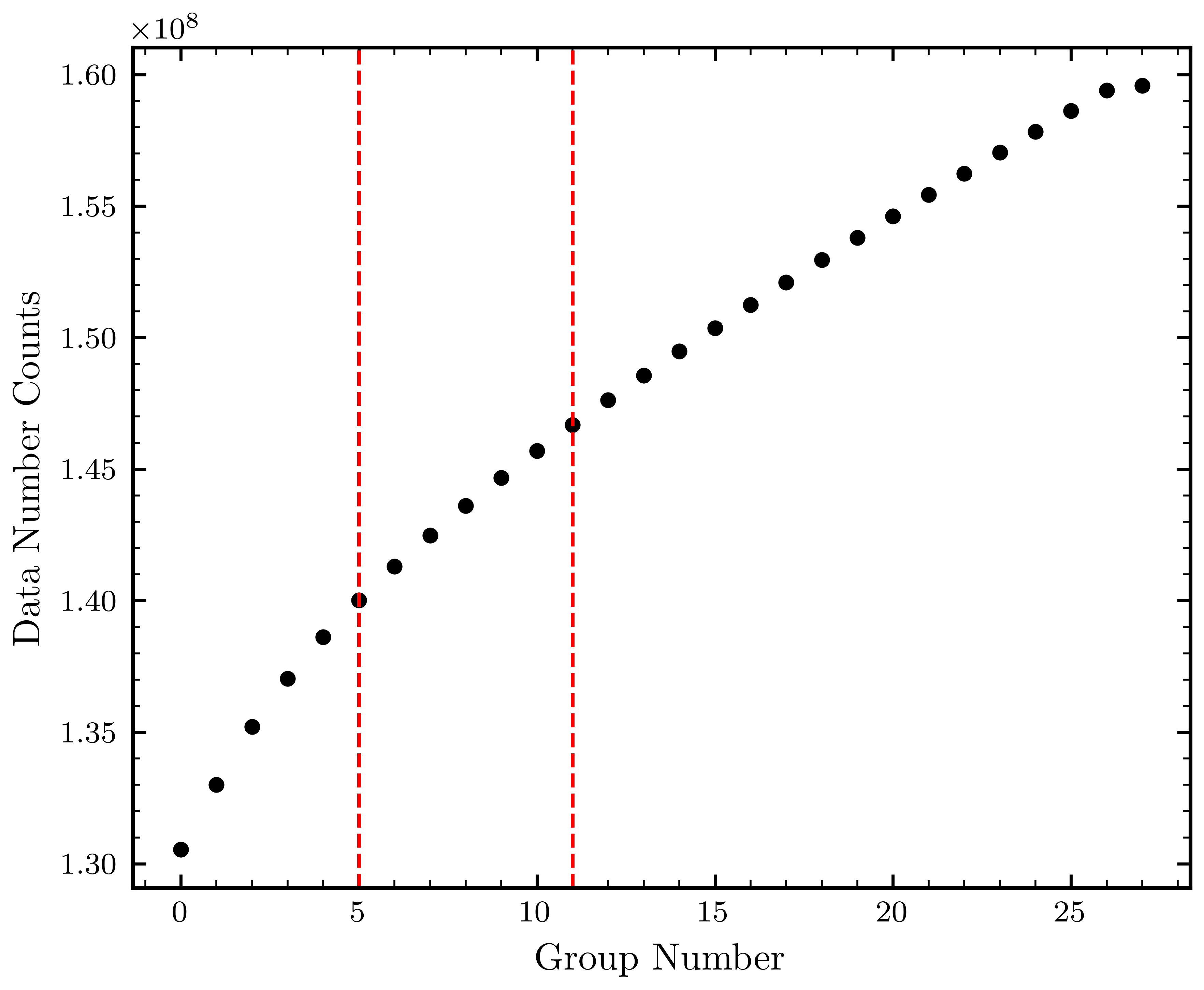} \\
\caption{Median data number counts summed over the data images as a function of group number after linearity correction of the first data segment. Non-linear behavior is seen in the first $\sim$11 groups, and a downward offset in the last group (\#28) is seen. Although each pixel behaves differently, this non-linearity is found in all integrations. Two red dashed lines indicate groups \#5 and \#11. The non-linearity could be caused by the reset-switch charge decay (RSCD; \citealt{morrison_jwst_2023, dyrek_transiting_2024}), which causes a greater increase in the ramp at the start of an integration. Besides our standard \texttt{Eureka!} reduction shown in \S\ref{sec:data_reduction}, we also implemented \texttt{RscdStep} (\url{https://jwst-pipeline.readthedocs.io/en/latest/api/jwst.rscd.RscdStep.html}) from the standard \texttt{jwst} pipeline in \texttt{Eureka!}. Similar to what our preferred \texttt{SPARTA} reduction did (i.e., excluding the first 5 groups), \texttt{RscdStep} excluded the first 4 groups from the up-the-ramp fitting for this dataset. The resulted best-fit white light eclipse depth is $147\pm$17\,ppm, compared to $143^{+17}_{-19}$\,ppm from our \texttt{Eureka!} standard reduction, $145\pm20$\,ppm from our preferred \texttt{SPARTA gr5} reduction and $141 \pm 17$\,ppm from the global fitting. 
\label{append_fig2}}
\end{figure}

\begin{figure}[ht]
\includegraphics[width = 9cm]{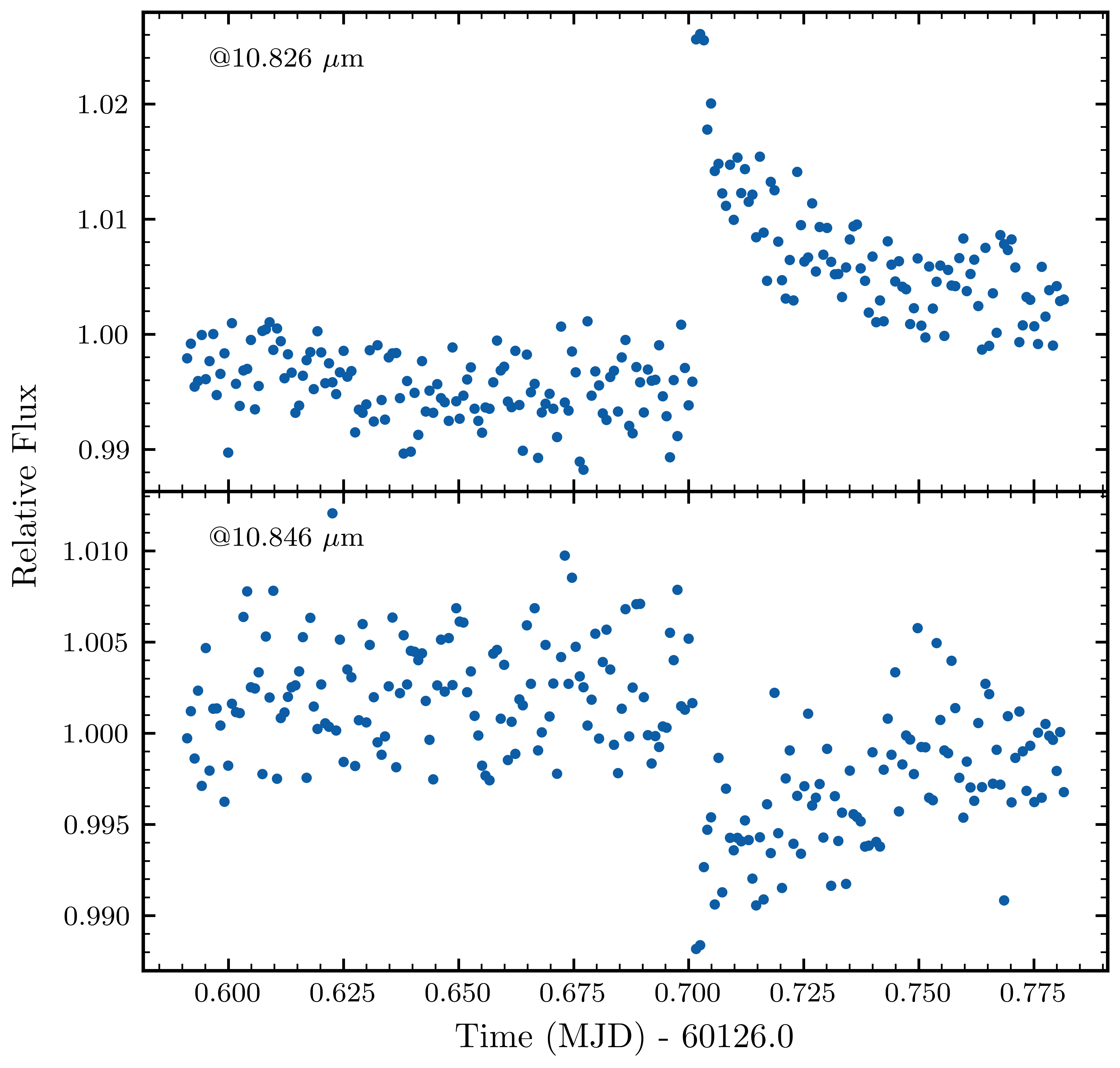} \\
\caption{Pixel-level light curves at 10.7861 and 10.8060 $\mu$m. An anomalous rise in flux followed by an exponential decay was found at 10.7861\,$\mu m$, and a sharp drop at 10.8060 $\mu$m was found at the same timestamp.
\label{append_fig4}}
\end{figure}

\pagebreak
\input{table1.tex}\label{app_tab}

\bibliographystyle{aasjournal}
\bibliography{reference}


\end{document}

%% file: table1.tex
\providecommand{\bjdtdb}{\ensuremath{\rm {BJD_{TDB}}}}
\providecommand{\tjdtdb}{\ensuremath{\rm {TJD_{TDB}}}}
\providecommand{\feh}{\ensuremath{\left[{\rm Fe}/{\rm H}\right]}}
\providecommand{\teff}{\ensuremath{T_{\rm eff}}}
\providecommand{\teq}{\ensuremath{T_{\rm eq}}}
\providecommand{\ecosw}{\ensuremath{e\cos{\omega_*}}}
\providecommand{\esinw}{\ensuremath{e\sin{\omega_*}}}
\providecommand{\msun}{\ensuremath{\,M_\Sun}}
\providecommand{\rsun}{\ensuremath{\,R_\Sun}}
\providecommand{\lsun}{\ensuremath{\,L_\Sun}}
\providecommand{\mj}{\ensuremath{\,M_{\rm J}}}
\providecommand{\rj}{\ensuremath{\,R_{\rm J}}}
\providecommand{\me}{\ensuremath{\,M_{\rm E}}}
\providecommand{\re}{\ensuremath{\,R_{\rm E}}}
\providecommand{\fave}{\langle F \rangle}
\providecommand{\fluxcgs}{10$^9$ erg s$^{-1}$ cm$^{-2}$}

\startlongtable

\begin{deluxetable*}{llc}
\tablecaption{Refined parameters for the GJ\,1132 system.}
\tablehead{\colhead{~~~Parameter} & \colhead{Description} & \multicolumn{1}{c}{Values}}
\startdata
\smallskip\\\multicolumn{1}{l}{Stellar Parameters:}&\smallskip\\
~~~~$M_*$\dotfill &Mass (\msun)\dotfill &$0.1945^{+0.0048}_{-0.0046}$\\
~~~~$R_*$\dotfill &Radius (\rsun)\dotfill &$0.2211^{+0.0069}_{-0.0081}$\\
~~~~$R_{*,SED}$\dotfill &Radius$^{1}$ (\rsun)\dotfill &$0.2257^{+0.0045}_{-0.0051}$\\
~~~~$L_*$\dotfill &Luminosity (\lsun)\dotfill &$0.00477^{+0.00036}_{-0.00026}$\\
~~~~$F_{Bol}$\dotfill &Bolometric Flux (cgs)\dotfill &$9.60^{+0.73}_{-0.53} \times 10^{-10}$\\
~~~~$\rho_*$\dotfill &Density (cgs)\dotfill &$25.3^{+3.1}_{-2.2}$\\
~~~~$\log{g}$\dotfill &Surface gravity (cgs)\dotfill &$5.037^{+0.034}_{-0.026}$\\
~~~~$T_{\rm eff}$\dotfill &Effective temperature (K)\dotfill &$3229^{+78}_{-62}$\\
~~~~$T_{\rm eff,SED}$\dotfill &Effective temperature$^{1}$ (K)\dotfill &$3197^{+69}_{-39}$\\
~~~~$[{\rm Fe/H}]$\dotfill &Metallicity (dex)\dotfill &$-0.17\pm0.15$\\
~~~~$K_S$\dotfill &Absolute Ks-band mag (mag)\dotfill &$7.817\pm0.027$\\
~~~~$k_S$\dotfill &Apparent Ks-band mag (mag)\dotfill &$8.320\pm0.027$\\
~~~~$A_V$\dotfill &V-band extinction (mag)\dotfill &$0.15^{+0.22}_{-0.11}$\\
~~~~$\sigma_{SED}$\dotfill &SED photometry error scaling \dotfill &$2.45^{+1.0}_{-0.63}$\\
~~~~$\varpi$\dotfill &Parallax (mas)\dotfill &$79.316\pm0.021$\\
~~~~$d$\dotfill &Distance (pc)\dotfill &$12.6078\pm0.0033$\\
\smallskip\\\multicolumn{1}{l}{Planetary Parameters:}&b\smallskip\\
~~~~$P$\dotfill &Period (days)\dotfill &$1.62892911^{+0.00000029}_{-0.00000030}$\\
~~~~$R_P$\dotfill &Radius (\rj)\dotfill &$0.1063^{+0.0034}_{-0.0040}$\\
~~~~$M_P$\dotfill &Mass (\mj)\dotfill &$0.00578^{+0.00060}_{-0.00059}$\\
~~~~$T_C$\dotfill &Observed Time of conjunction$^{2}$ (\bjdtdb)\dotfill &$2459280.98988\pm0.00013$\\
~~~~$T_C$\dotfill &Model Time of conjunction$^{2,3}$ (\tjdtdb)\dotfill &$2459280.98979\pm0.00013$\\
~~~~$T_T$\dotfill &Model time of min proj sep$^{3,4,5}$ (\tjdtdb)\dotfill &$2459994.460742^{+0.000018}_{-0.000019}$\\
~~~~$T_0$\dotfill &Obs time of min proj sep$^{4,6,7}$ (\bjdtdb)\dotfill &$2459994.460834\pm0.000018$\\
~~~~$a$\dotfill &Semi-major axis (AU)\dotfill &$0.01570\pm0.00013$\\
~~~~$i$\dotfill &Inclination (Degrees)\dotfill &$88.16^{+0.19}_{-0.18}$\\
~~~~$e$\dotfill &Eccentricity \dotfill &$0.0118^{+0.047}_{-0.0099}$\\
~~~~$\omega_*$\dotfill &Argument of periastron (Degrees)\dotfill &$-95.8^{+5.5}_{-130}$\\
~~~~$\dot{\omega}_{\rm GR}$\dotfill &Computed GR precession ($^\circ$/century)\dotfill &$2.966^{+0.050}_{-0.048}$\\
~~~~$T_{\rm eq}$\dotfill &Equilibrium temp$^{8}$ (K)\dotfill &$583.8^{+11}_{-8.5}$\\
~~~~$\tau_{\rm circ}$\dotfill &Tidal circularization timescale (Gyr)\dotfill &$3.56^{+0.75}_{-0.59}$\\
~~~~$K$\dotfill &RV semi-amplitude (m/s)\dotfill &$2.98\pm0.30$\\
~~~~$R_P/R_*$\dotfill &Radius of planet in stellar radii \dotfill &$0.04943\pm0.00015$\\
~~~~$a/R_*$\dotfill &Semi-major axis in stellar radii \dotfill &$15.26^{+0.59}_{-0.45}$\\
~~~~$\delta$\dotfill &$\left(R_P/R_*\right)^2$ \dotfill &$0.002443^{+0.000014}_{-0.000015}$\\
~~~~$\delta_{\rm MIRILRS}$\dotfill &Transit depth in MIRI LRS (frac)\dotfill &$0.002477^{+0.000054}_{-0.000030}$\\
~~~~$\delta_{\rm TESS}$\dotfill &Transit depth in TESS (frac)\dotfill &$0.00261^{+0.00025}_{-0.00012}$\\
~~~~$\tau$\dotfill &Transit in/egress duration (days)\dotfill &$0.001989^{+0.000095}_{-0.000097}$\\
~~~~$T_{14}$\dotfill &Total transit duration (days)\dotfill &$0.031936^{+0.000089}_{-0.000090}$\\
~~~~$T_{FWHM}$\dotfill &FWHM transit duration (days)\dotfill &$0.029946^{+0.000057}_{-0.000056}$\\
~~~~$b$\dotfill &Transit impact parameter \dotfill &$0.504^{+0.032}_{-0.038}$\\
~~~~$b_S$\dotfill &Eclipse impact parameter \dotfill &$0.481^{+0.039}_{-0.042}$\\
~~~~$\tau_S$\dotfill &In/egress eclipse duration (days)\dotfill &$0.00189^{+0.00014}_{-0.00018}$\\
~~~~$T_{S,14}$\dotfill &Total eclipse duration (days)\dotfill &$0.03151^{+0.00051}_{-0.0022}$\\
~~~~$T_{S,FWHM}$\dotfill &FWHM eclipse duration (days)\dotfill &$0.02958^{+0.00044}_{-0.0020}$\\
~~~~$\delta_{S,2.5\mu m}$\dotfill &BB eclipse depth at 2.5$\mu$m (ppm)\dotfill &$0.635^{+0.10}_{-0.078}$\\
~~~~$\delta_{S,5.0\mu m}$\dotfill &BB eclipse depth at 5.0$\mu$m (ppm)\dotfill &$25.7^{+1.9}_{-1.6}$\\
~~~~$\delta_{S,7.5\mu m}$\dotfill &BB eclipse depth at 7.5$\mu$m (ppm)\dotfill &$77.2^{+3.6}_{-3.4}$\\
~~~~$\rho_P$\dotfill &Density (cgs)\dotfill &$5.97^{+0.96}_{-0.79}$\\
~~~~$logg_P$\dotfill &Surface gravity (cgs)\dotfill &$3.104\pm0.054$\\
~~~~$\Theta$\dotfill &Safronov Number \dotfill &$0.00878^{+0.00097}_{-0.00093}$\\
~~~~$\fave$\dotfill &Incident Flux (\fluxcgs)\dotfill &$0.0263^{+0.0020}_{-0.0015}$\\
~~~~$T_S$\dotfill &Observed Time of eclipse$^{2}$ (\bjdtdb)\dotfill &$2459281.80312^{+0.00088}_{-0.00070}$\\
~~~~$T_S$\dotfill &Model Time of eclipse$^{2,3}$ (\tjdtdb)\dotfill &$2459281.80321^{+0.00087}_{-0.00070}$\\
~~~~$T_E$\dotfill &Model time of sec min proj sep$^{3,4,5}$ (\tjdtdb)\dotfill &$2460128.84633^{+0.00087}_{-0.00067}$\\
~~~~$T_{E,0}$\dotfill &Obs time of sec min proj sep$^{4,6,7}$ (\bjdtdb)\dotfill &$2460128.84624^{+0.00087}_{-0.00067}$\\
~~~~$T_P$\dotfill &Time of Periastron (\tjdtdb)\dotfill &$2459280.149^{+0.025}_{-0.60}$\\
~~~~$T_A$\dotfill &Time of ascending node (\tjdtdb)\dotfill &$2459280.5772^{+0.0056}_{-0.025}$\\
~~~~$T_D$\dotfill &Time of descending node (\tjdtdb)\dotfill &$2459281.4011^{+0.026}_{-0.0055}$\\
~~~~$V_c/V_e$\dotfill &Scaled velocity \dotfill &$1.009^{+0.051}_{-0.011}$\\
~~~~$e\cos{\omega_*}$\dotfill & \dotfill &$-0.00102^{+0.00084}_{-0.00065}$\\
~~~~$e\sin{\omega_*}$\dotfill & \dotfill &$-0.009^{+0.011}_{-0.050}$\\
~~~~$M_P\sin i$\dotfill &Minimum mass (\mj)\dotfill &$0.00578^{+0.00060}_{-0.00059}$\\
~~~~$M_P/M_*$\dotfill &Mass ratio \dotfill &$2.84\pm0.29 \times 10^{-5}$\\
~~~~$d/R_*$\dotfill &Separation at mid transit \dotfill &$15.42^{+1.3}_{-0.60}$\\
~~~~$P_T$\dotfill &A priori non-grazing transit probability \dotfill &$0.0616^{+0.0025}_{-0.0048}$\\
~~~~$P_{T,G}$\dotfill &A priori transit probability \dotfill &$0.0680^{+0.0028}_{-0.0053}$\\
~~~~$P_S$\dotfill &A priori non-grazing eclipse probability \dotfill &$0.0636^{+0.0014}_{-0.0015}$\\
~~~~$P_{S,G}$\dotfill &A priori eclipse prob \dotfill &$0.0702\pm0.0016$\\
\smallskip\\\multicolumn{2}{l}{Wavelength Parameters:}\\
~~MIRI LRS\smallskip\\
~~~~$u_{1}$\dotfill &Linear limb-darkening coeff \dotfill          & $0.039^{+0.054}_{-0.029}$ \\
~~~~$u_{2}$\dotfill &Quadratic limb-darkening coeff \dotfill       & $0.178^{+0.051}_{-0.075}$ \\
~~~~$A_T$\dotfill &Thermal emission from the planet (ppm)\dotfill  & $140\pm17$ \\
~~~~$\delta_{S}$\dotfill &Measured eclipse depth (ppm)\dotfill     & $140\pm17$ \\ \smallskip
~~TESS\smallskip\\
~~~~$u_{1}$\dotfill &Linear limb-darkening coeff \dotfill          & $0.18^{+0.22}_{-0.13}$ \\
~~~~$u_{2}$\dotfill &Quadratic limb-darkening coeff \dotfill       & $0.04^{+0.19}_{-0.15}$ \\
~~~~$A_T$\dotfill &Thermal emission from the planet (ppm)\dotfill  & -- \\
~~~~$\delta_{S}$\dotfill &Measured eclipse depth (ppm)\dotfill     & -- \\
                             &   \\
Telescope Parameters          &   \\
~~HARPS \smallskip \\
~~~~$\gamma_{\rm rel}$\dotfill &Relative RV Offset (m/s)\dotfill &$35078.84\pm0.25$\\
~~~~$\sigma_J$\dotfill &RV Jitter (m/s)\dotfill &$0.49\pm0.49$\\
~~~~$\sigma_J^2$\dotfill &RV Jitter Variance \dotfill &$0.24^{+0.73}_{-0.59}$\\
                             &   \\
Transit Parameters            &   \\
TESS UT 2019-03-01 (TESS) \smallskip    &   \\
~~~~$\sigma^{2}$\dotfill &Added Variance \dotfill & $-1.61\pm0.40 \times 10^{-7}$ \\
~~~~$\tau_{Ramp}$\dotfill &Exponential Ramp \dotfill& -- \\
~~~~$A_{Ramp}$\dotfill &Amplitude of exp ramp \dotfill & -- \\
~~~~$F_0$\dotfill &Baseline flux \dotfill & $1.000042\pm0.000015$ \\
~~~~$C_{0}$\dotfill &Additive detrending coeff \dotfill & -- \\
~~~~$C_{1}$\dotfill &Additive detrending coeff \dotfill& -- \\
JWST UT 2023-02-25 (MIRILRS) &   \\
~~~~$\sigma^{2}$\dotfill &Added Variance \dotfill & $-1.1734^{+0.0020}_{-0.0018} \times 10^{-6}$\\
~~~~$\tau_{Ramp}$\dotfill &Exponential Ramp \dotfill& $0.50^{+0.35}_{-0.32}$\\
~~~~$A_{Ramp}$\dotfill &Amp of exp ramp \dotfill& $0.00180^{+0.0011}_{-0.00098}$\\
~~~~$F_0$\dotfill &Baseline flux \dotfill &$1.000141^{+0.000026}_{-0.000024}$\\
~~~~$C_{0}$\dotfill &Additive detrending coeff \dotfill & -- \\
~~~~$C_{1}$\dotfill &Additive detrending coeff \dotfill& -- \\
JWST UT 2023-02-25 (MIRILRS) &   \\
~~~~$\sigma^{2}$\dotfill &Added Variance \dotfill &$-1.0451^{+0.0031}_{-0.0029} \times 10^{-6}$\\
~~~~$\tau_{Ramp}$\dotfill &Exponential Ramp \dotfill& $0.70^{+0.21}_{-0.32}$\\
~~~~$A_{Ramp}$\dotfill &Amp of exp ramp \dotfill& $-0.00057^{+0.00026}_{-0.00029}$\\
~~~~$F_0$\dotfill &Baseline flux \dotfill &$0.999877\pm0.000025$ \\
~~~~$C_{0}$\dotfill &Additive detrending coeff \dotfill & -- \\
~~~~$C_{1}$\dotfill &Additive detrending coeff\dotfill& -- \\
JWST UT 2023-03-05 (MIRILRS) &   \\
~~~~$\sigma^{2}$\dotfill &Added Variance \dotfill  & $-1.0882^{+0.0021}_{-0.0019} \times 10^{-6}$\\
~~~~$\tau_{Ramp}$\dotfill &Exponential Ramp \dotfill&$0.72^{+0.20}_{-0.29}$\\
~~~~$A_{Ramp}$\dotfill &Amp of exp ramp \dotfill&$0.00124^{+0.00037}_{-0.00045}$\\
~~~~$F_0$\dotfill &Baseline flux \dotfill &$1.000043\pm0.000023$ \\
~~~~$C_{0}$\dotfill &Additive detrending coeff \dotfill & -- \\
~~~~$C_{1}$\dotfill &Additive detrending coeff\dotfill& -- \\
JWST UT 2023-03-05 (MIRILRS) &   \\
~~~~$\sigma^{2}$\dotfill &Added Variance \dotfill &$-1.0261^{+0.0031}_{-0.0029} \times 10^{-6}$ \\
~~~~$\tau_{Ramp}$\dotfill &Exponential Ramp \dotfill&$0.70^{+0.22}_{-0.33}$\\
~~~~$A_{Ramp}$\dotfill &Amp of exp ramp\dotfill &$-0.00036^{+0.00020}_{-0.00025}$\\
~~~~$F_0$\dotfill &Baseline flux \dotfill &$0.999910\pm0.000025$ \\
~~~~$C_{0}$\dotfill &Additive detrending coeff \dotfill & -- \\
~~~~$C_{1}$\dotfill &Additive detrending coeff\dotfill&-- \\
JWST UT 2023-07-01 (MIRILRS) &   \\
~~~~$\sigma^{2}$\dotfill &Added Variance \dotfill & $-7.2^{+2.8}_{-2.6} \times 10^{-9}$ \\
~~~~$\tau_{Ramp}$\dotfill &Exponential Ramp \dotfill &$0.0305^{+0.012}_{-0.0081}$\\
~~~~$A_{Ramp}$\dotfill &Amp of exp ramp \dotfill&$0.000249^{+0.000032}_{-0.000030}$\\
~~~~$F_0$\dotfill &Baseline flux \dotfill &$1.000220\pm0.000038$\\
~~~~$C_{0}$\dotfill &Additive detrending coeff \dotfill  & $-0.000402^{+0.000027}_{-0.000026}$ \\
~~~~$C_{1}$\dotfill &Additive detrending coeff\dotfill& $-0.000150^{+0.000026}_{-0.000027}$\\
\enddata
\label{tab:GJ1132.}
\tablenotetext{}{See Table 3 in \citet{Eastman:2019} for a detailed description of all parameters}
\tablenotetext{1}{This value ignores the systematic error and is for reference only}
\tablenotetext{2}{Time of conjunction is commonly reported as the ``transit time''}
\tablenotetext{3}{\tjdtdb is the target's barycentric frame and corrects for light travel time}
\tablenotetext{4}{Time of minimum projected separation is a more correct ``transit time''}
\tablenotetext{5}{Use this to model TTVs, e}
\tablenotetext{6}{At the epoch that minimizes the covariance between $T_C$ and Period}
\tablenotetext{7}{Use this to predict future transit times}
\tablenotetext{8}{Assumes no albedo and perfect redistribution}
\end{deluxetable*}


%% file: manuscript.bbl
\begin{thebibliography}{}
\expandafter\ifx\csname natexlab\endcsname\relax\def\natexlab#1{#1}\fi
\providecommand{\url}[1]{\href{#1}{#1}}
\providecommand{\dodoi}[1]{doi:~\href{http://doi.org/#1}{\nolinkurl{#1}}}
\providecommand{\doeprint}[1]{\href{http://ascl.net/#1}{\nolinkurl{http://ascl.net/#1}}}
\providecommand{\doarXiv}[1]{\href{https://arxiv.org/abs/#1}{\nolinkurl{https://arxiv.org/abs/#1}}}

\bibitem[{Affolter {et~al.}(2023)Affolter, Mordasini, Oza, Kubyshkina, \& Fossati}]{affolter_planetary_2023}
Affolter, L., Mordasini, C., Oza, A.~V., Kubyshkina, D., \& Fossati, L. 2023, Astronomy \& Astrophysics, 676, A119, \dodoi{10.1051/0004-6361/202142205}

\bibitem[{Alderson {et~al.}(2024)Alderson, Batalha, Wakeford, Wallack, Aguichine, Teske, Redai, Alam, Batalha, Gao, Kirk, Lopez-Morales, Moran, Scarsdale, Wogan, \& Wolfgang}]{alderson_jwst_2024}
Alderson, L., Batalha, N.~E., Wakeford, H.~R., {et~al.} 2024, {JWST} {COMPASS}: {NIRSpec}/{G395H} {Transmission} {Observations} of the {Super}-{Earth} {TOI}-836b,  arXiv.
\newblock \url{http://arxiv.org/abs/2404.00093}

\bibitem[{{Allard} {et~al.}(2012){Allard}, {Homeier}, \& {Freytag}}]{phoenix}
{Allard}, F., {Homeier}, D., \& {Freytag}, B. 2012, Philosophical Transactions of the Royal Society of London Series A, 370, 2765, \dodoi{10.1098/rsta.2011.0269}

\bibitem[{{Astropy Collaboration} {et~al.}(2013){Astropy Collaboration}, {Robitaille}, {Tollerud}, {Greenfield}, {Droettboom}, {Bray}, {Aldcroft}, {Davis}, {Ginsburg}, {Price-Whelan}, {Kerzendorf}, {Conley}, {Crighton}, {Barbary}, {Muna}, {Ferguson}, {Grollier}, {Parikh}, {Nair}, {Unther}, {Deil}, {Woillez}, {Conseil}, {Kramer}, {Turner}, {Singer}, {Fox}, {Weaver}, {Zabalza}, {Edwards}, {Azalee Bostroem}, {Burke}, {Casey}, {Crawford}, {Dencheva}, {Ely}, {Jenness}, {Labrie}, {Lim}, {Pierfederici}, {Pontzen}, {Ptak}, {Refsdal}, {Servillat}, \& {Streicher}}]{astropy:2013}
{Astropy Collaboration}, {Robitaille}, T.~P., {Tollerud}, E.~J., {et~al.} 2013, \aap, 558, A33, \dodoi{10.1051/0004-6361/201322068}

\bibitem[{{Astropy Collaboration} {et~al.}(2018){Astropy Collaboration}, {Price-Whelan}, {Sip{\H{o}}cz}, {G{\"u}nther}, {Lim}, {Crawford}, {Conseil}, {Shupe}, {Craig}, {Dencheva}, {Ginsburg}, {Vand erPlas}, {Bradley}, {P{\'e}rez-Su{\'a}rez}, {de Val-Borro}, {Aldcroft}, {Cruz}, {Robitaille}, {Tollerud}, {Ardelean}, {Babej}, {Bach}, {Bachetti}, {Bakanov}, {Bamford}, {Barentsen}, {Barmby}, {Baumbach}, {Berry}, {Biscani}, {Boquien}, {Bostroem}, {Bouma}, {Brammer}, {Bray}, {Breytenbach}, {Buddelmeijer}, {Burke}, {Calderone}, {Cano Rodr{\'\i}guez}, {Cara}, {Cardoso}, {Cheedella}, {Copin}, {Corrales}, {Crichton}, {D'Avella}, {Deil}, {Depagne}, {Dietrich}, {Donath}, {Droettboom}, {Earl}, {Erben}, {Fabbro}, {Ferreira}, {Finethy}, {Fox}, {Garrison}, {Gibbons}, {Goldstein}, {Gommers}, {Greco}, {Greenfield}, {Groener}, {Grollier}, {Hagen}, {Hirst}, {Homeier}, {Horton}, {Hosseinzadeh}, {Hu}, {Hunkeler}, {Ivezi{\'c}}, {Jain}, {Jenness}, {Kanarek}, {Kendrew}, {Kern}, {Kerzendorf}, {Khvalko}, {King}, {Kirkby}, {Kulkarni},
  {Kumar}, {Lee}, {Lenz}, {Littlefair}, {Ma}, {Macleod}, {Mastropietro}, {McCully}, {Montagnac}, {Morris}, {Mueller}, {Mumford}, {Muna}, {Murphy}, {Nelson}, {Nguyen}, {Ninan}, {N{\"o}the}, {Ogaz}, {Oh}, {Parejko}, {Parley}, {Pascual}, {Patil}, {Patil}, {Plunkett}, {Prochaska}, {Rastogi}, {Reddy Janga}, {Sabater}, {Sakurikar}, {Seifert}, {Sherbert}, {Sherwood-Taylor}, {Shih}, {Sick}, {Silbiger}, {Singanamalla}, {Singer}, {Sladen}, {Sooley}, {Sornarajah}, {Streicher}, {Teuben}, {Thomas}, {Tremblay}, {Turner}, {Terr{\'o}n}, {van Kerkwijk}, {de la Vega}, {Watkins}, {Weaver}, {Whitmore}, {Woillez}, {Zabalza}, \& {Astropy Contributors}}]{astropy:2018}
{Astropy Collaboration}, {Price-Whelan}, A.~M., {Sip{\H{o}}cz}, B.~M., {et~al.} 2018, \aj, 156, 123, \dodoi{10.3847/1538-3881/aabc4f}

\bibitem[{{Astropy Collaboration} {et~al.}(2022){Astropy Collaboration}, {Price-Whelan}, {Lim}, {Earl}, {Starkman}, {Bradley}, {Shupe}, {Patil}, {Corrales}, {Brasseur}, {N{"o}the}, {Donath}, {Tollerud}, {Morris}, {Ginsburg}, {Vaher}, {Weaver}, {Tocknell}, {Jamieson}, {van Kerkwijk}, {Robitaille}, {Merry}, {Bachetti}, {G{"u}nther}, {Aldcroft}, {Alvarado-Montes}, {Archibald}, {B{'o}di}, {Bapat}, {Barentsen}, {Baz{'a}n}, {Biswas}, {Boquien}, {Burke}, {Cara}, {Cara}, {Conroy}, {Conseil}, {Craig}, {Cross}, {Cruz}, {D'Eugenio}, {Dencheva}, {Devillepoix}, {Dietrich}, {Eigenbrot}, {Erben}, {Ferreira}, {Foreman-Mackey}, {Fox}, {Freij}, {Garg}, {Geda}, {Glattly}, {Gondhalekar}, {Gordon}, {Grant}, {Greenfield}, {Groener}, {Guest}, {Gurovich}, {Handberg}, {Hart}, {Hatfield-Dodds}, {Homeier}, {Hosseinzadeh}, {Jenness}, {Jones}, {Joseph}, {Kalmbach}, {Karamehmetoglu}, {Ka{l}uszy{'n}ski}, {Kelley}, {Kern}, {Kerzendorf}, {Koch}, {Kulumani}, {Lee}, {Ly}, {Ma}, {MacBride}, {Maljaars}, {Muna}, {Murphy}, {Norman}, {O'Steen},
  {Oman}, {Pacifici}, {Pascual}, {Pascual-Granado}, {Patil}, {Perren}, {Pickering}, {Rastogi}, {Roulston}, {Ryan}, {Rykoff}, {Sabater}, {Sakurikar}, {Salgado}, {Sanghi}, {Saunders}, {Savchenko}, {Schwardt}, {Seifert-Eckert}, {Shih}, {Jain}, {Shukla}, {Sick}, {Simpson}, {Singanamalla}, {Singer}, {Singhal}, {Sinha}, {Sip{H{o}}cz}, {Spitler}, {Stansby}, {Streicher}, {{{S}}umak}, {Swinbank}, {Taranu}, {Tewary}, {Tremblay}, {Val-Borro}, {Van Kooten}, {Vasovi{'c}}, {Verma}, {de Miranda Cardoso}, {Williams}, {Wilson}, {Winkel}, {Wood-Vasey}, {Xue}, {Yoachim}, {Zhang}, {Zonca}, \& {Astropy Project Contributors}}]{astropy:2022}
{Astropy Collaboration}, {Price-Whelan}, A.~M., {Lim}, P.~L., {et~al.} 2022, \apj, 935, 167, \dodoi{10.3847/1538-4357/ac7c74}

\bibitem[{Barnes {et~al.}(2018)Barnes, Deitrick, Luger, Driscoll, Quinn, Fleming, Guyer, McDonald, Meadows, Arney, Crisp, Domagal-Goldman, Foreman-Mackey, Kaib, Lincowski, Lustig-Yaeger, \& Schwieterman}]{barnes_habitability_2018}
Barnes, R., Deitrick, R., Luger, R., {et~al.} 2018, The {Habitability} of {Proxima} {Centauri} b {I}: {Evolutionary} {Scenarios},  arXiv.
\newblock \url{http://arxiv.org/abs/1608.06919}

\bibitem[{Bell {et~al.}(2022)Bell, Ahrer, Brande, Carter, Feinstein, Guzman~Caloca, Mansfield, Zieba, Piaulet, Benneke, Filippazzo, May, Roy, Kreidberg, \& Stevenson}]{bell_eureka_2022}
Bell, T.~J., Ahrer, E.-M., Brande, J., {et~al.} 2022, Journal of Open Source Software, 7, 4503, \dodoi{10.21105/joss.04503}

\bibitem[{{Bell} {et~al.}(2024){Bell}, {Crouzet}, {Cubillos}, {Kreidberg}, {Piette}, {Roman}, {Barstow}, {Blecic}, {Carone}, {Coulombe}, {Ducrot}, {Hammond}, {Mendon{\c{c}}a}, {Moses}, {Parmentier}, {Stevenson}, {Teinturier}, {Zhang}, {Batalha}, {Bean}, {Benneke}, {Charnay}, {Chubb}, {Demory}, {Gao}, {Lee}, {L{\'o}pez-Morales}, {Morello}, {Rauscher}, {Sing}, {Tan}, {Venot}, {Wakeford}, {Aggarwal}, {Ahrer}, {Alam}, {Baeyens}, {Barrado}, {Caceres}, {Carter}, {Casewell}, {Challener}, {Crossfield}, {Decin}, {D{\'e}sert}, {Dobbs-Dixon}, {Dyrek}, {Espinoza}, {Feinstein}, {Gibson}, {Harrington}, {Helling}, {Hu}, {Iro}, {Kempton}, {Kendrew}, {Komacek}, {Krick}, {Lagage}, {Leconte}, {Lendl}, {Lewis}, {Lothringer}, {Malsky}, {Mancini}, {Mansfield}, {Mayne}, {Mikal-Evans}, {Molaverdikhani}, {Nikolov}, {Nixon}, {Palle}, {Petit dit de la Roche}, {Piaulet}, {Powell}, {Rackham}, {Schneider}, {Steinrueck}, {Taylor}, {Welbanks}, {Yurchenko}, {Zhang}, \& {Zieba}}]{bell_w43_2024}
{Bell}, T.~J., {Crouzet}, N., {Cubillos}, P.~E., {et~al.} 2024, arXiv e-prints, arXiv:2401.13027, \dodoi{10.48550/arXiv.2401.13027}

\bibitem[{Berta-Thompson {et~al.}(2015)Berta-Thompson, Irwin, Charbonneau, Newton, Dittmann, Astudillo-Defru, Bonfils, Gillon, Jehin, Stark, Stalder, Bouchy, Delfosse, Forveille, Lovis, Mayor, Neves, Pepe, Santos, Udry, \& Wünsche}]{berta-thompson_rocky_2015}
Berta-Thompson, Z.~K., Irwin, J., Charbonneau, D., {et~al.} 2015, Nature, 527, 204, \dodoi{10.1038/nature15762}

\bibitem[{Bonfils {et~al.}(2018)Bonfils, Almenara, Cloutier, Wünsche, Astudillo-Defru, Berta-Thompson, Bouchy, Charbonneau, Delfosse, Díaz, Dittmann, Doyon, Forveille, Irwin, Lovis, Mayor, Menou, Murgas, Newton, Pepe, Santos, \& Udry}]{bonfils_radial_2018}
Bonfils, X., Almenara, J.-M., Cloutier, R., {et~al.} 2018, Astronomy \& Astrophysics, 618, A142, \dodoi{10.1051/0004-6361/201731884}

\bibitem[{{Bouwman} {et~al.}(2023){Bouwman}, {Kendrew}, {Greene}, {Bell}, {Lagage}, {Schreiber}, {Dicken}, {Sloan}, {Espinoza}, {Scheithauer}, {Coulais}, {Fox}, {Gastaud}, {Glauser}, {Jones}, {Labiano}, {Lahuis}, {Morrison}, {Murray}, {Mueller}, {Nayak}, {Wright}, {Glasse}, \& {Rieke}}]{bouwman_miri_2023}
{Bouwman}, J., {Kendrew}, S., {Greene}, T.~P., {et~al.} 2023, \pasp, 135, 038002, \dodoi{10.1088/1538-3873/acbc49}

\bibitem[{Bushouse {et~al.}(2022)Bushouse, Eisenhamer, Dencheva, Davies, Greenfield, Morrison, Hodge, Simon, Grumm, Droettboom, Slavich, Sosey, Pauly, Miller, Jedrzejewski, Hack, Davis, Crawford, Law, Gordon, Regan, Cara, MacDonald, Bradley, Shanahan, Jamieson, Teodoro, \& Williams}]{bushouse_2022_7325378}
Bushouse, H., Eisenhamer, J., Dencheva, N., {et~al.} 2022, JWST Calibration Pipeline, 1.8.2,  Zenodo, \dodoi{10.5281/zenodo.7325378}

\bibitem[{{Cadieux} {et~al.}(2024){Cadieux}, {Doyon}, {MacDonald}, {Turbet}, {Artigau}, {Lim}, {Radica}, {Fauchez}, {Salhi}, {Dang}, {Albert}, {Coulombe}, {Cowan}, {Lafreni{\`e}re}, {L'Heureux}, {Piaulet}, {Benneke}, {Cloutier}, {Charnay}, {Cook}, {Fournier-Tondreau}, {Plotnykov}, \& {Valencia}}]{cadieux24_lhs1140}
{Cadieux}, C., {Doyon}, R., {MacDonald}, R.~J., {et~al.} 2024, arXiv e-prints, arXiv:2406.15136, \dodoi{10.48550/arXiv.2406.15136}

\bibitem[{Cowan \& Agol(2011)}]{cowan_statistics_2011}
Cowan, N.~B., \& Agol, E. 2011, The Astrophysical Journal, 729, 54, \dodoi{10.1088/0004-637X/729/1/54}

\bibitem[{{Crossfield} {et~al.}(2022){Crossfield}, {Malik}, {Hill}, {Kane}, {Foley}, {Polanski}, {Coria}, {Brande}, {Zhang}, {Wienke}, {Kreidberg}, {Cowan}, {Dragomir}, {Gorjian}, {Mikal-Evans}, {Benneke}, {Christiansen}, {Deming}, \& {Morales}}]{crossfield_gj_2022}
{Crossfield}, I. J.~M., {Malik}, M., {Hill}, M.~L., {et~al.} 2022, \apjl, 937, L17, \dodoi{10.3847/2041-8213/ac886b}

\bibitem[{Deming {et~al.}(2009)Deming, Seager, Winn, Miller-Ricci, Clampin, Lindler, Greene, Charbonneau, Laughlin, Ricker, Latham, \& Ennico}]{deming_discovery_2009}
Deming, D., Seager, S., Winn, J., {et~al.} 2009, Publications of the Astronomical Society of the Pacific, 121, 952, \dodoi{10.1086/605913}

\bibitem[{{Dittmann} {et~al.}(2017){Dittmann}, {Irwin}, {Charbonneau}, {Berta-Thompson}, \& {Newton}}]{dittmann_search_2017}
{Dittmann}, J.~A., {Irwin}, J.~M., {Charbonneau}, D., {Berta-Thompson}, Z.~K., \& {Newton}, E.~R. 2017, \aj, 154, 142, \dodoi{10.3847/1538-3881/aa855b}

\bibitem[{Do~Amaral {et~al.}(2022)Do~Amaral, Barnes, Segura, \& Luger}]{do_amaral_contribution_2022}
Do~Amaral, L. N.~R., Barnes, R., Segura, A., \& Luger, R. 2022, The Astrophysical Journal, 928, 12, \dodoi{10.3847/1538-4357/ac53af}

\bibitem[{Dressing \& Charbonneau(2015)}]{dressing_occurrence_2015}
Dressing, C.~D., \& Charbonneau, D. 2015, The Astrophysical Journal, 807, 45, \dodoi{10.1088/0004-637X/807/1/45}

\bibitem[{Dyrek {et~al.}(2024)Dyrek, Ducrot, Lagage, Tremblin, Kendrew, Bouwman, \& Bouffet}]{dyrek_transiting_2024}
Dyrek, A., Ducrot, E., Lagage, P.-O., {et~al.} 2024, Astronomy \& Astrophysics, 683, A212, \dodoi{10.1051/0004-6361/202347127}

\bibitem[{{Eastman} {et~al.}(2019){Eastman}, {Rodriguez}, {Agol}, {Stassun}, {Beatty}, {Vanderburg}, {Gaudi}, {Collins}, \& {Luger}}]{Eastman:2019}
{Eastman}, J.~D., {Rodriguez}, J.~E., {Agol}, E., {et~al.} 2019, arXiv e-prints, arXiv:1907.09480.
\newblock \doarXiv{1907.09480}

\bibitem[{{Emery} {et~al.}(1998){Emery}, {Sprague}, {Witteborn}, {Colwell}, {Kozlowski}, \& {Wooden}}]{emery_1988}
{Emery}, J.~P., {Sprague}, A.~L., {Witteborn}, F.~C., {et~al.} 1998, \icarus, 136, 104, \dodoi{10.1006/icar.1998.6012}

\bibitem[{Foreman-Mackey {et~al.}(2012)Foreman-Mackey, Hogg, Lang, \& Goodman}]{foreman-mackey_emcee_2012}
Foreman-Mackey, D., Hogg, D.~W., Lang, D., \& Goodman, J. 2012, \dodoi{10.48550/ARXIV.1202.3665}

\bibitem[{Greene {et~al.}(2023)Greene, Bell, Ducrot, Dyrek, Lagage, \& Fortney}]{greene_thermal_2023}
Greene, T.~P., Bell, T.~J., Ducrot, E., {et~al.} 2023, Nature, 618, 39, \dodoi{10.1038/s41586-023-05951-7}

\bibitem[{Harris {et~al.}(2020)Harris, Millman, van~der Walt, Gommers, Virtanen, Cournapeau, Wieser, Taylor, Berg, Smith, Kern, Picus, Hoyer, van Kerkwijk, Brett, Haldane, del R{\'{i}}o, Wiebe, Peterson, G{\'{e}}rard-Marchant, Sheppard, Reddy, Weckesser, Abbasi, Gohlke, \& Oliphant}]{harris2020array}
Harris, C.~R., Millman, K.~J., van~der Walt, S.~J., {et~al.} 2020, Nature, 585, 357, \dodoi{10.1038/s41586-020-2649-2}

\bibitem[{Henry {et~al.}(2018)Henry, Jao, Winters, Dieterich, Finch, Ianna, Riedel, Silverstein, Subasavage, \& Vrijmoet}]{henry_solar_2018}
Henry, T.~J., Jao, W.-C., Winters, J.~G., {et~al.} 2018, The Astronomical Journal, 155, 265, \dodoi{10.3847/1538-3881/aac262}

\bibitem[{{Hu} {et~al.}(2012){Hu}, {Ehlmann}, \& {Seager}}]{hu2012}
{Hu}, R., {Ehlmann}, B.~L., \& {Seager}, S. 2012, \apj, 752, 7, \dodoi{10.1088/0004-637X/752/1/7}

\bibitem[{Hunter(2007)}]{Hunter:2007}
Hunter, J.~D. 2007, Computing in Science \& Engineering, 9, 90, \dodoi{10.1109/MCSE.2007.55}

\bibitem[{{Ih} {et~al.}(2023){Ih}, {Kempton}, {Whittaker}, \& {Lessard}}]{Ih_T1b_2023}
{Ih}, J., {Kempton}, E. M.~R., {Whittaker}, E.~A., \& {Lessard}, M. 2023, \apjl, 952, L4, \dodoi{10.3847/2041-8213/ace03b}

\bibitem[{Iyer {et~al.}(2023)Iyer, Line, Muirhead, Fortney, \& Gharib-Nezhad}]{iyer_sphinx_2023}
Iyer, A.~R., Line, M.~R., Muirhead, P.~S., Fortney, J.~J., \& Gharib-Nezhad, E. 2023, The Astrophysical Journal, 944, 41, \dodoi{10.3847/1538-4357/acabc2}

\bibitem[{Kempton {et~al.}(2018)Kempton, Bean, Louie, Deming, Koll, Mansfield, Christiansen, López-Morales, Swain, Zellem, Ballard, Barclay, Barstow, Batalha, Beatty, Berta-Thompson, Birkby, Buchhave, Charbonneau, Cowan, Crossfield, Val-Borro, Doyon, Dragomir, Gaidos, Heng, Hu, Kane, Kreidberg, Mallonn, Morley, Narita, Nascimbeni, Pallé, Quintana, Rauscher, Seager, Shkolnik, Sing, Sozzetti, Stassun, Valenti, \& Essen}]{kempton_framework_2018}
Kempton, E. M.-R., Bean, J.~L., Louie, D.~R., {et~al.} 2018, Publications of the Astronomical Society of the Pacific, 130, 114401, \dodoi{10.1088/1538-3873/aadf6f}

\bibitem[{Kempton {et~al.}(2023)Kempton, Zhang, Bean, Steinrueck, Piette, Parmentier, Malsky, Roman, Rauscher, Gao, Bell, Xue, Taylor, Savel, Arnold, Nixon, Stevenson, Mansfield, Kendrew, Zieba, Ducrot, Dyrek, Lagage, Stassun, Henry, Barman, Lupu, Malik, Kataria, Ih, Fu, Welbanks, \& McGill}]{kempton_reflective_2023}
Kempton, E. M.-R., Zhang, M., Bean, J.~L., {et~al.} 2023, Nature, 620, 67, \dodoi{10.1038/s41586-023-06159-5}

\bibitem[{{Kendrew} {et~al.}(2015){Kendrew}, {Scheithauer}, {Bouchet}, {Amiaux}, {Azzollini}, {Bouwman}, {Chen}, {Dubreuil}, {Fischer}, {Glasse}, {Greene}, {Lagage}, {Lahuis}, {Ronayette}, {Wright}, \& {Wright}}]{kendrew_mid-infrared_2015}
{Kendrew}, S., {Scheithauer}, S., {Bouchet}, P., {et~al.} 2015, \pasp, 127, 623, \dodoi{10.1086/682255}

\bibitem[{Kirk {et~al.}(2024)Kirk, Stevenson, Fu, Lustig-Yaeger, Moran, Peacock, Alam, Batalha, Bennett, Gonzalez-Quiles, López-Morales, Lothringer, MacDonald, May, Mayorga, Rustamkulov, Sing, Sotzen, Valenti, \& Wakeford}]{kirk_jwstnircam_2024}
Kirk, J., Stevenson, K.~B., Fu, G., {et~al.} 2024, The Astronomical Journal, 167, 90, \dodoi{10.3847/1538-3881/ad19df}

\bibitem[{Kite \& Barnett(2020)}]{kite_exoplanet_2020}
Kite, E.~S., \& Barnett, M.~N. 2020, Proceedings of the National Academy of Sciences, 117, 18264, \dodoi{10.1073/pnas.2006177117}

\bibitem[{{Kite} \& {Schaefer}(2021)}]{kite_schaefer2021}
{Kite}, E.~S., \& {Schaefer}, L. 2021, \apjl, 909, L22, \dodoi{10.3847/2041-8213/abe7dc}

\bibitem[{{Koll}(2022)}]{koll_scaling_2022}
{Koll}, D. D.~B. 2022, \apj, 924, 134, \dodoi{10.3847/1538-4357/ac3b48}

\bibitem[{{Koll} {et~al.}(2019){Koll}, {Malik}, {Mansfield}, {Kempton}, {Kite}, {Abbot}, \& {Bean}}]{koll_identifying_2019}
{Koll}, D. D.~B., {Malik}, M., {Mansfield}, M., {et~al.} 2019, \apj, 886, 140, \dodoi{10.3847/1538-4357/ab4c91}

\bibitem[{Kreidberg(2015)}]{kreidberg_batman_2015}
Kreidberg, L. 2015, Publications of the Astronomical Society of the Pacific, 127, 1161, \dodoi{10.1086/683602}

\bibitem[{{Kreidberg} {et~al.}(2019){Kreidberg}, {Koll}, {Morley}, {Hu}, {Schaefer}, {Deming}, {Stevenson}, {Dittmann}, {Vanderburg}, {Berardo}, {Guo}, {Stassun}, {Crossfield}, {Charbonneau}, {Latham}, {Loeb}, {Ricker}, {Seager}, \& {Vanderspek}}]{kreidberg_absence_2019}
{Kreidberg}, L., {Koll}, D. D.~B., {Morley}, C., {et~al.} 2019, \nat, 573, 87, \dodoi{10.1038/s41586-019-1497-4}

\bibitem[{{Li} {et~al.}(2021){Li}, {Bergin}, {Blake}, {Ciesla}, \& {Hirschmann}}]{li_carbon_2021}
{Li}, J., {Bergin}, E.~A., {Blake}, G.~A., {Ciesla}, F.~J., \& {Hirschmann}, M.~M. 2021, Science Advances, 7, eabd3632, \dodoi{10.1126/sciadv.abd3632}

\bibitem[{Libby-Roberts {et~al.}(2022)Libby-Roberts, Berta-Thompson, Diamond-Lowe, Gully-Santiago, Irwin, Kempton, Rackham, Charbonneau, Désert, Dittmann, Hofmann, Morley, \& Newton}]{libby-roberts_featureless_2022}
Libby-Roberts, J.~E., Berta-Thompson, Z.~K., Diamond-Lowe, H., {et~al.} 2022, The Astronomical Journal, 164, 59, \dodoi{10.3847/1538-3881/ac75de}

\bibitem[{Lim {et~al.}(2023)Lim, Benneke, Doyon, MacDonald, Piaulet, Artigau, Coulombe, Radica, L’Heureux, Albert, Rackham, De~Wit, Salhi, Roy, Flagg, Fournier-Tondreau, Taylor, Cook, Lafrenière, Cowan, Kaltenegger, Rowe, Espinoza, Dang, \& Darveau-Bernier}]{lim_atmospheric_2023}
Lim, O., Benneke, B., Doyon, R., {et~al.} 2023, The Astrophysical Journal Letters, 955, L22, \dodoi{10.3847/2041-8213/acf7c4}

\bibitem[{{Lincowski} {et~al.}(2023){Lincowski}, {Meadows}, {Zieba}, {Kreidberg}, {Morley}, {Gillon}, {Selsis}, {Agol}, {Bolmont}, {Ducrot}, {Hu}, {Koll}, {Lyu}, {Mandell}, {Suissa}, \& {Tamburo}}]{lincowski2023}
{Lincowski}, A.~P., {Meadows}, V.~S., {Zieba}, S., {et~al.} 2023, \apjl, 955, L7, \dodoi{10.3847/2041-8213/acee02}

\bibitem[{Luger \& Barnes(2015)}]{luger_extreme_2015}
Luger, R., \& Barnes, R. 2015, Astrobiology, 15, 119, \dodoi{10.1089/ast.2014.1231}

\bibitem[{Lustig-Yaeger {et~al.}(2023)Lustig-Yaeger, Fu, May, Ceballos, Moran, Peacock, Stevenson, Kirk, López-Morales, MacDonald, Mayorga, Sing, Sotzen, Valenti, Redai, Alam, Batalha, Bennett, Gonzalez-Quiles, Kruse, Lothringer, Rustamkulov, \& Wakeford}]{lustig-yaeger_jwst_2023}
Lustig-Yaeger, J., Fu, G., May, E.~M., {et~al.} 2023, Nature Astronomy, 7, 1317, \dodoi{10.1038/s41550-023-02064-z}

\bibitem[{{Lyu} {et~al.}(2024){Lyu}, {Koll}, {Cowan}, {Hu}, {Kreidberg}, \& {Rose}}]{lyu2024}
{Lyu}, X., {Koll}, D. D.~B., {Cowan}, N.~B., {et~al.} 2024, \apj, 964, 152, \dodoi{10.3847/1538-4357/ad2077}

\bibitem[{Lyu {et~al.}(2024)Lyu, Koll, Cowan, Hu, Kreidberg, \& Rose}]{lyu_super-earth_2024}
Lyu, X., Koll, D. D.~B., Cowan, N.~B., {et~al.} 2024, Super-{Earth} {LHS3844b} is tidally locked,  arXiv.
\newblock \url{http://arxiv.org/abs/2310.01725}

\bibitem[{{Mahajan} {et~al.}(2024){Mahajan}, {Eastman}, \& {Kirk}}]{Mahajan:2024}
{Mahajan}, A.~S., {Eastman}, J.~D., \& {Kirk}, J. 2024, \apjl, 963, L37, \dodoi{10.3847/2041-8213/ad29f3}

\bibitem[{{Malik} {et~al.}(2019{\natexlab{a}}){Malik}, {Kempton}, {Koll}, {Mansfield}, {Bean}, \& {Kite}}]{helios3}
{Malik}, M., {Kempton}, E. M.~R., {Koll}, D. D.~B., {et~al.} 2019{\natexlab{a}}, \apj, 886, 142, \dodoi{10.3847/1538-4357/ab4a05}

\bibitem[{{Malik} {et~al.}(2019{\natexlab{b}}){Malik}, {Kitzmann}, {Mendon{\c{c}}a}, {Grimm}, {Marleau}, {Linder}, {Tsai}, \& {Heng}}]{helios2}
{Malik}, M., {Kitzmann}, D., {Mendon{\c{c}}a}, J.~M., {et~al.} 2019{\natexlab{b}}, \aj, 157, 170, \dodoi{10.3847/1538-3881/ab1084}

\bibitem[{{Malik} {et~al.}(2017){Malik}, {Grosheintz}, {Mendon{\c{c}}a}, {Grimm}, {Lavie}, {Kitzmann}, {Tsai}, {Burrows}, {Kreidberg}, {Bedell}, {Bean}, {Stevenson}, \& {Heng}}]{helios1}
{Malik}, M., {Grosheintz}, L., {Mendon{\c{c}}a}, J.~M., {et~al.} 2017, \aj, 153, 56, \dodoi{10.3847/1538-3881/153/2/56}

\bibitem[{{Mansfield} {et~al.}(2019){Mansfield}, {Kite}, {Hu}, {Koll}, {Malik}, {Bean}, \& {Kempton}}]{mansfield_identifying_2019}
{Mansfield}, M., {Kite}, E.~S., {Hu}, R., {et~al.} 2019, \apj, 886, 141, \dodoi{10.3847/1538-4357/ab4c90}

\bibitem[{May {et~al.}(2023)May, MacDonald, Bennett, Moran, Wakeford, Peacock, Lustig-Yaeger, Highland, Stevenson, Sing, Mayorga, Batalha, Kirk, López-Morales, Valenti, Alam, Alderson, Fu, Gonzalez-Quiles, Lothringer, Rustamkulov, \& Sotzen}]{may_double_2023}
May, E.~M., MacDonald, R.~J., Bennett, K.~A., {et~al.} 2023, The Astrophysical Journal, 959, L9, \dodoi{10.3847/2041-8213/ad054f}

\bibitem[{Moran {et~al.}(2023)Moran, Stevenson, Sing, MacDonald, Kirk, Lustig-Yaeger, Peacock, Mayorga, Bennett, López-Morales, May, Rustamkulov, Valenti, Adams~Redai, Alam, Batalha, Fu, Gonzalez-Quiles, Highland, Kruse, Lothringer, Ortiz~Ceballos, Sotzen, \& Wakeford}]{moran_high_2023}
Moran, S.~E., Stevenson, K.~B., Sing, D.~K., {et~al.} 2023, The Astrophysical Journal Letters, 948, L11, \dodoi{10.3847/2041-8213/accb9c}

\bibitem[{Morrison {et~al.}(2023)Morrison, Dicken, Argyriou, Ressler, Gordon, Regan, Cracraft, Rieke, Engesser, Alberts, Alvarez-Marquez, Colbert, Fox, Gasman, Law, Garcia~Marin, Gáspár, Guillard, Kendrew, Labiano, Laine, Noriega-Crespo, Shivaei, \& Sloan}]{morrison_jwst_2023}
Morrison, J.~E., Dicken, D., Argyriou, I., {et~al.} 2023, Publications of the Astronomical Society of the Pacific, 135, 075004, \dodoi{10.1088/1538-3873/acdea6}

\bibitem[{Mugnai {et~al.}(2021)Mugnai, Modirrousta-Galian, Edwards, Changeat, Bouwman, Morello, Al-Refaie, Baeyens, Bieger, Blain, Gressier, Guilluy, Jaziri, Kiefer, Morvan, Pluriel, Poveda, Skaf, Whiteford, Wright, Yip, Zingales, Charnay, Drossart, Leconte, Venot, Waldmann, \& Beaulieu}]{mugnai_ares_2021}
Mugnai, L.~V., Modirrousta-Galian, D., Edwards, B., {et~al.} 2021, The Astronomical Journal, 161, 284, \dodoi{10.3847/1538-3881/abf3c3}

\bibitem[{Mulders {et~al.}(2015)Mulders, Pascucci, \& Apai}]{mulders_stellar-mass-dependent_2015}
Mulders, G.~D., Pascucci, I., \& Apai, D. 2015, The Astrophysical Journal, 798, 112, \dodoi{10.1088/0004-637X/798/2/112}

\bibitem[{Nakayama {et~al.}(2022)Nakayama, Ikoma, \& Terada}]{nakayama_survival_2022}
Nakayama, A., Ikoma, M., \& Terada, N. 2022, The Astrophysical Journal, 937, 72, \dodoi{10.3847/1538-4357/ac86ca}

\bibitem[{{Powell} {et~al.}(2024){Powell}, {Feinstein}, {Lee}, {Zhang}, {Tsai}, {Taylor}, {Kirk}, {Bell}, {Barstow}, {Gao}, {Bean}, {Blecic}, {Chubb}, {Crossfield}, {Jordan}, {Kitzmann}, {Moran}, {Morello}, {Moses}, {Welbanks}, {Yang}, {Zhang}, {Ahrer}, {Bello-Arufe}, {Brande}, {Casewell}, {Crouzet}, {Cubillos}, {Demory}, {Dyrek}, {Flagg}, {Hu}, {Inglis}, {Jones}, {Kreidberg}, {L{\'o}pez-Morales}, {Lagage}, {Meier Vald{\'e}s}, {Miguel}, {Parmentier}, {Piette}, {Rackham}, {Radica}, {Redfield}, {Stevenson}, {Wakeford}, {Aggarwal}, {Alam}, {Batalha}, {Batalha}, {Benneke}, {Berta-Thompson}, {Brady}, {Caceres}, {Carter}, {D{\'e}sert}, {Harrington}, {Iro}, {Line}, {Lothringer}, {MacDonald}, {Mancini}, {Molaverdikhani}, {Mukherjee}, {Nixon}, {Oza}, {Palle}, {Rustamkulov}, {Sing}, {Steinrueck}, {Venot}, {Wheatley}, \& {Yurchenko}}]{powell_sulfur_2024}
{Powell}, D., {Feinstein}, A.~D., {Lee}, E. K.~H., {et~al.} 2024, \nat, 626, 979, \dodoi{10.1038/s41586-024-07040-9}

\bibitem[{{Redfield} {et~al.}(2024){Redfield}, {Batalha}, {Benneke}, {Biller}, {Espinoza}, {France}, {Konopacky}, {Kreidberg}, {Rauscher}, \& {Sing}}]{redfield24}
{Redfield}, S., {Batalha}, N., {Benneke}, B., {et~al.} 2024, arXiv e-prints, arXiv:2404.02932, \dodoi{10.48550/arXiv.2404.02932}

\bibitem[{Reylé {et~al.}(2021)Reylé, Jardine, Fouqué, Caballero, Smart, \& Sozzetti}]{reyle_10_2021}
Reylé, C., Jardine, K., Fouqué, P., {et~al.} 2021, Astronomy \& Astrophysics, 650, A201, \dodoi{10.1051/0004-6361/202140985}

\bibitem[{Ribas {et~al.}(2016)Ribas, Bolmont, Selsis, Reiners, Leconte, Raymond, Engle, Guinan, Morin, Turbet, Forget, \& Anglada-Escudé}]{ribas_habitability_2016}
Ribas, I., Bolmont, E., Selsis, F., {et~al.} 2016, Astronomy \& Astrophysics, 596, A111, \dodoi{10.1051/0004-6361/201629576}

\bibitem[{Sabotta {et~al.}(2021)Sabotta, Schlecker, Chaturvedi, Guenther, Muñoz~Rodríguez, Muñoz~Sánchez, Caballero, Shan, Reffert, Ribas, Reiners, Hatzes, Amado, Klahr, Morales, Quirrenbach, Henning, Dreizler, Pallé, Perger, Azzaro, Jeffers, Kaminski, Kürster, Lafarga, Montes, Passegger, \& Zechmeister}]{sabotta_carmenes_2021}
Sabotta, S., Schlecker, M., Chaturvedi, P., {et~al.} 2021, Astronomy \& Astrophysics, 653, A114, \dodoi{10.1051/0004-6361/202140968}

\bibitem[{Schlawin {et~al.}(2023)Schlawin, Beatty, Brooks, Nikolov, Greene, Espinoza, Glidic, Baka, Egami, Stansberry, Boyer, Gennaro, Leisenring, Hilbert, Misselt, Kelly, Canipe, Beichman, Correnti, Knight, Jurling, Perrin, Feinberg, McElwain, Bond, Ciardi, Kendrew, \& Rieke}]{schlawin_jwst_2023}
Schlawin, E., Beatty, T., Brooks, B., {et~al.} 2023, Publications of the Astronomical Society of the Pacific, 135, 018001, \dodoi{10.1088/1538-3873/aca718}

\bibitem[{Segura {et~al.}(2010)Segura, Walkowicz, Meadows, Kasting, \& Hawley}]{segura_effect_2010}
Segura, A., Walkowicz, L.~M., Meadows, V., Kasting, J., \& Hawley, S. 2010, Astrobiology, 10, 751, \dodoi{10.1089/ast.2009.0376}

\bibitem[{Selsis {et~al.}(2011)Selsis, Wordsworth, \& Forget}]{selsis_thermal_2011}
Selsis, F., Wordsworth, R.~D., \& Forget, F. 2011, Astronomy \& Astrophysics, 532, A1, \dodoi{10.1051/0004-6361/201116654}

\bibitem[{Speagle(2019)}]{speagle_dynesty_2019}
Speagle, J.~S. 2019, \dodoi{10.48550/ARXIV.1904.02180}

\bibitem[{{Spencer}(1990)}]{spencer_1990}
{Spencer}, J.~R. 1990, \icarus, 83, 27, \dodoi{10.1016/0019-1035(90)90004-S}

\bibitem[{{STScI Development Team}(2013)}]{pysynphot}
{STScI Development Team}. 2013, {pysynphot: Synthetic photometry software package}, Astrophysics Source Code Library, record ascl:1303.023

\bibitem[{Swain {et~al.}(2021)Swain, Estrela, Roudier, Sotin, Rimmer, Valio, West, Pearson, Huber-Feely, \& Zellem}]{swain_detection_2021}
Swain, M.~R., Estrela, R., Roudier, G.~M., {et~al.} 2021, The Astronomical Journal, 161, 213, \dodoi{10.3847/1538-3881/abe879}

\bibitem[{Vidotto {et~al.}(2013)Vidotto, Jardine, Morin, Donati, Lang, \& Russell}]{vidotto_effects_2013}
Vidotto, A.~A., Jardine, M., Morin, J., {et~al.} 2013, Astronomy \& Astrophysics, 557, A67, \dodoi{10.1051/0004-6361/201321504}

\bibitem[{Virtanen {et~al.}(2020)Virtanen, Gommers, Oliphant, Haberland, Reddy, Cournapeau, Burovski, Peterson, Weckesser, Bright, {van der Walt}, Brett, Wilson, Millman, Mayorov, Nelson, Jones, Kern, Larson, Carey, Polat, Feng, Moore, {VanderPlas}, Laxalde, Perktold, Cimrman, Henriksen, Quintero, Harris, Archibald, Ribeiro, Pedregosa, {van Mulbregt}, \& {SciPy 1.0 Contributors}}]{2020SciPy-NMeth}
Virtanen, P., Gommers, R., Oliphant, T.~E., {et~al.} 2020, Nature Methods, 17, 261, \dodoi{10.1038/s41592-019-0686-2}

\bibitem[{West {et~al.}(2008)West, Hawley, Bochanski, Covey, Reid, Dhital, Hilton, \& Masuda}]{west_constraining_2008}
West, A.~A., Hawley, S.~L., Bochanski, J.~J., {et~al.} 2008, The Astronomical Journal, 135, 785, \dodoi{10.1088/0004-6256/135/3/785}

\bibitem[{{Whittaker} {et~al.}(2022){Whittaker}, {Malik}, {Ih}, {Kempton}, {Mansfield}, {Bean}, {Kite}, {Koll}, {Cronin}, \& {Hu}}]{Whittaker_lhs3844_2022}
{Whittaker}, E.~A., {Malik}, M., {Ih}, J., {et~al.} 2022, \aj, 164, 258, \dodoi{10.3847/1538-3881/ac9ab3}

\bibitem[{Xue {et~al.}(2024{\natexlab{a}})Xue, Bean, Zhang, Welbanks, Lunine, \& August}]{xue_jwst_2024}
Xue, Q., Bean, J.~L., Zhang, M., {et~al.} 2024{\natexlab{a}}, The Astrophysical Journal Letters, 963, L5, \dodoi{10.3847/2041-8213/ad2682}

\bibitem[{Xue {et~al.}(2024{\natexlab{b}})Xue, Bean, Zhang, Mahajan, Ih, Eastman, Lunine, Weiner~Mansfield, Park~Coy, M.-R.~Kempton, Koll, \& Kite}]{xue_2024_13244543}
Xue, Q., Bean, J., Zhang, M., {et~al.} 2024{\natexlab{b}}, {Data and codes for 'JWST Thermal Emission of the Terrestrial Exoplanet GJ 1132b'},  Zenodo, \dodoi{10.5281/zenodo.13244543}

\bibitem[{Zahnle \& Catling(2017)}]{zahnle_cosmic_2017}
Zahnle, K.~J., \& Catling, D.~C. 2017, The Astrophysical Journal, 843, 122, \dodoi{10.3847/1538-4357/aa7846}

\bibitem[{Zendejas {et~al.}(2010)Zendejas, Segura, \& Raga}]{zendejas_atmospheric_2010}
Zendejas, J., Segura, A., \& Raga, A. 2010, Icarus, 210, 539, \dodoi{10.1016/j.icarus.2010.07.013}

\bibitem[{Zhang {et~al.}(2024)Zhang, Hu, Inglis, Dai, Bean, Knutson, Lam, Goffo, \& Gandolfi}]{zhang_gj_2024}
Zhang, M., Hu, R., Inglis, J., {et~al.} 2024, The Astrophysical Journal Letters, 961, L44, \dodoi{10.3847/2041-8213/ad1a07}

\bibitem[{Zieba {et~al.}(2023)Zieba, Kreidberg, Ducrot, Gillon, Morley, Schaefer, Tamburo, Koll, Lyu, Acuña, Agol, Iyer, Hu, Lincowski, Meadows, Selsis, Bolmont, Mandell, \& Suissa}]{zieba_no_2023}
Zieba, S., Kreidberg, L., Ducrot, E., {et~al.} 2023, Nature, 620, 746, \dodoi{10.1038/s41586-023-06232-z}

\end{thebibliography}
